\documentclass[%
reprint,
superscriptaddress,
amsmath,amssymb,
aps,
]{revtex4-2}

\usepackage{graphicx}
\usepackage{dcolumn}
\usepackage{bm,array}
\usepackage{hyperref}
\usepackage{cleveref}
\usepackage{comment}

\usepackage{subfigure}
\usepackage{xcolor}

\begin{document}

\preprint{APS/123-QED}

\title{Fatigue failure in glasses under cyclic shear deformation}


\author{Swarnendu Maity}
\affiliation{Jawaharlal Nehru Center for Advanced Scientific Research, Jakkur Campus, Bengaluru 560064, India.}

\author{Himangsu Bhaumik}
\affiliation{Jawaharlal Nehru Center for Advanced Scientific Research, Jakkur Campus, Bengaluru 560064, India.}
\affiliation{Yusuf Hamied Department of Chemistry, University of Cambridge, Lensfield Road, Cambridge CB2 1EW, UK}
\author{Shivakumar Athani}%
\affiliation{Jawaharlal Nehru Center for Advanced Scientific Research, Jakkur Campus, Bengaluru 560064, India.}

\author{Srikanth Sastry}
\affiliation{Jawaharlal Nehru Center for Advanced Scientific Research, Jakkur Campus, Bengaluru 560064, India.}

\date{\today}


\begin{abstract}
Solids subjected to repeated cycles of stress or deformation can fail after several cycles, a phenomenon termed fatigue failure. Although intensely investigated for a wide range of materials owing to its obvious practical importance, a microscopic understanding of the initiation of fatigue failure continues to be actively pursued, in particular for soft and amorphous materials. We investigate fatigue failure for glasses subjected to cyclic shear deformation through computer simulations. We show that, approaching the so-called fatigue limit, failure times display a power law divergence, at variance with commonly used functional forms, and exhibit strong dependence on the degree of annealing of the glasses. We explore several measures of {\it damage}, based on quantification of plastic rearrangements and on dissipated energy. Strikingly, the fraction of particles that undergo plastic rearrangements, and a percolation transition they undergo, are predictive of failure. We also find a robust power law relationship between accumulated damage, quantified by dissipated energy or non-affine displacements, and the failure times, which permits prediction of failure times based on behaviour in the initial cycles. These observations reveal salient new microscopic features of fatigue failure and suggest approaches for developing a full microscopic picture of fatigue failure in amorphous solids. 
\end{abstract}


\maketitle

The mechanical response of a solid to applied stress, or deformation, is a material property of obvious importance. Such response for large enough applied stresses involves plastic rearrangements and eventually leads to yielding and failure. Of particular importance is the phenomenon of fatigue failure, wherein a solid, subjected to repeated cycles of stress, or deformation, fails after several cycles of stress or deformation \cite{SureshCUP1998,Christensen2014}. The number of cycles to failure, or the failure time, increases upon reduction of stress, or deformation, towards a threshold value termed the fatigue or endurance limit. Below the fatigue limit, a solid can be subjected to an indefinite number of cycles of loading without failing. Several empirical forms have been used to describe the variation of failure times upon approaching the fatigue limit. Descriptions of fatigue failure, developed over decades for metallic, crystalline materials, are expressed in terms of the formation and growth of microcracks, and correspondingly, the accumulation of {\it damage}, which is quantified by several measurable properties \cite{NaderiProcRSocA2010,JMKrishnan_2023,Bhowmik_2022,BhowmikPRE2022}. The initiation of failure at the microscopic level occurs before macroscopic signatures are detectable, but is essential to understand in developing a theoretical explanation of the fatigue failure process. The characterisation of such initiation, especially for soft and amorphous solids, is elusive \cite{KunJStat2007,LeomachPRL14,aime2018microscopic,Cipelletti2020,Bhowmik_2022,BhowmikPRE2022}, and is closely related to an understanding of yielding and creep behaviour in these materials. The disorder inherent to amorphous materials, and the corresponding heterogeneity of response, exhibit a rich diversity and  pose challenges to understanding their mechanical response, which have been investigated actively in recent years \cite{BonnRevModPhys2017,NicolasRevModPhys2018,Kumar_2024,SollichArxiv2024,LernerJNonCrys2019}. 

In particular, theoretical and computational investigation of yielding behaviour under cyclic shear deformation has been actively investigated recently\cite{FioccoPRE2013,Regev2013,Priezjev2013,FioccoIOP2015,LeishanthemNatCom2017, kawasakiPRE16,ParmarPRX2019,BhaumikPNAS2021,YehPRL2020,BhaumikPNAS2021,SastryPRL2021,MunganPRL2021,LiuJCP2022,ParleyPRL2022,CochranPRL2024,PRIEZJEV2023112230,CochranPRL2024}. Upon cyclic shear, with increasing applied strain amplitude, glasses undergo a transition marked 
(in the steady states reached after repeated cyclic deformation) by a change from non-diffusive to diffusive change of particle (cycle to cycle) positions \cite{FioccoPRE2013,Regev2013,FioccoIOP2015,LeishanthemNatCom2017,kawasakiPRE16,ParmarPRX2019,BhaumikPNAS2021}, and shear banding above the transition. The nature of the transition also displays a striking dependence on the degree of annealing of the glasses, with a qualitative change in behavior across a threshold degree of annealing, that intriguiingly appears  to be related to the mode coupling crossover of glassy dynamics \cite{BhaumikPNAS2021}.
The number of cycles to reach steady states appears to diverge as the transition point is  approached from either side. The dependence, on the distance from the transition (which we identify as the fatigue limit), of the number of cycles to failure is an important characteristic of the phenomenon of fatigue failure, which we investigate in detail for a model glass in the present work. 


\begin{figure*}[!htb]
\centerline{ \includegraphics[width=0.98\textwidth]{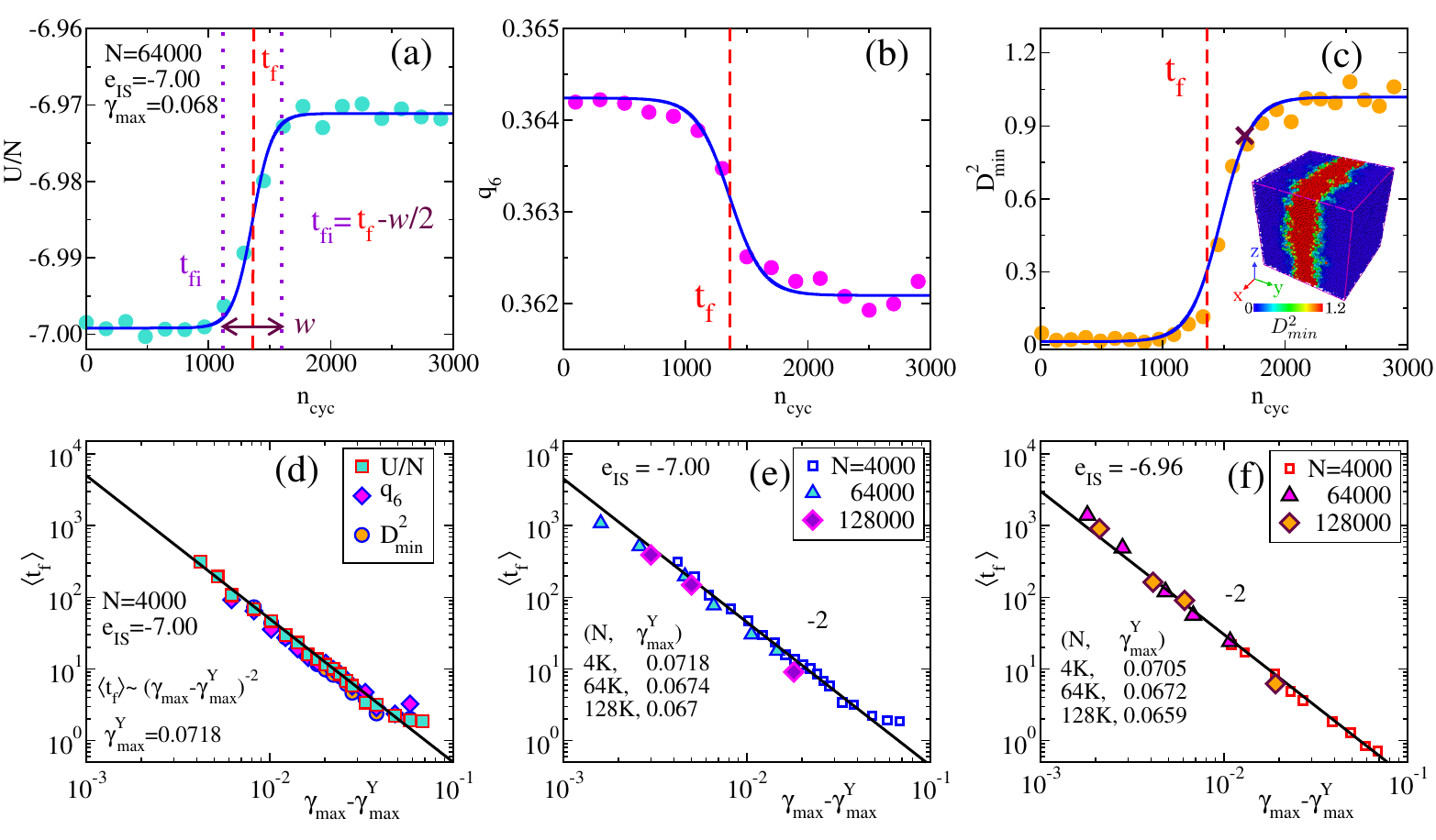}}
    \caption{ \textbf{Failure time and its divergence.} Variation of (a) Potential energy per particle, $U/N$, (b) Bond orientational parameter $q_6$, and (c) Non-affine displacement $D^2_{min}$ as a function of strain cycle $n_{cyc}$ for $\gamma_{max}=0.068$ and $N=64000$. Solid lines in (a)-(c) are fits to a sigmoid function, whose inflection point is identified as the failure time ($t_f\approx 1360$ for the case shown) indicated by the vertical dashed line. Dotted vertical lines in (a) indicate the width $w$ of the time interval over which failure occurs along with the time of initiation of failure $t_{fi}=t_f-w/2$. Inset in (c) shows a snap shot of the system just after the failure at $n_{cyc}=1670$ (marked by the cross in the main panel). (d) Sample-averaged failure times $\langle t_f \rangle $, estimated from the three different metrics ($U/N$ ,$q_6$, and $D_{min}^2$), as a function of $\gamma_{max}-\gamma^Y_{max}$ exhibit a power-law divergence with exponent $-2$. $\langle t_f \rangle$ as a function of $\gamma_{max}-\gamma_{max}^Y$ for different system sizes for two different degrees of annealing (e) $e_{IS}=-7.00$ and (f) $e_{IS}=-6.96$. The yield strain amplitudes for different system sizes are indicated in the figure. Lines indicate the power law with exponent $-2$. $t_f$ in (e,f) is estimated from $U/N$ data.
    }
\label{fig:failure_time_calc_eIS-7.00_diffmetric}
\end{figure*}

We perform cyclic shear simulations at a constant shear rate, at a very low temperature, for a large number of samples of a model glass, for different system sizes and annealing, and a range of shear deformation amplitudes, including amplitudes for which the glasses exhibit failure. The details of the simulations are provided in Methods, and in the Supplementary information (SI). Our results show that failure times diverge in a power law fashion at the fatigue limit, and show strong dependence on the degree of annealing of the glasses. We explore the relationship between different cumulative measures of plasticity and find striking correlations with failure times. Particles undergoing plastic rearrangements accumulate in a spatially heterogeneous manner and reach a nearly fixed fraction at the time of failure, and percolate as a precursor to shear band formation. Accumulated energy dissipation, as an alternate measure of plasticity, exhibits a power law relationship with failure times, and the near constancy of energy dissipation per cycle permits prediction of failure times from the energy dissipated in the initial cycles.  We present these results below, and close with a discussion of their implications.

\noindent{\bf {\em Identification of the yield strain amplitude/fatigue limit:}}
Upon application of repeated cycles of strain, previous work has shown that glasses approach limits states wherein the cycle-to-cycle displacements of particles either vanish (for small strain amplitude $\gamma_{max}$) or 
reach finite steady state values (for large enough strain $\gamma_{max}>\gamma_{max}^Y$, $\gamma_{max}^Y$ being the yield strain amplitude) and exhibit diffusive behaviour  \cite{FioccoPRE2013, LeishanthemNatCom2017}. We use the average per cycle displacements  of particles ($\Delta r$) or equivalently the mean squared displacements ($MSD$) to identify the yield strain $\gamma_{max}^Y$, as shown in the SI (Sec:\ref{SI_gammaY}). 

\begin{figure*}[ht!]
    \centering
    \includegraphics[width=0.95\linewidth]{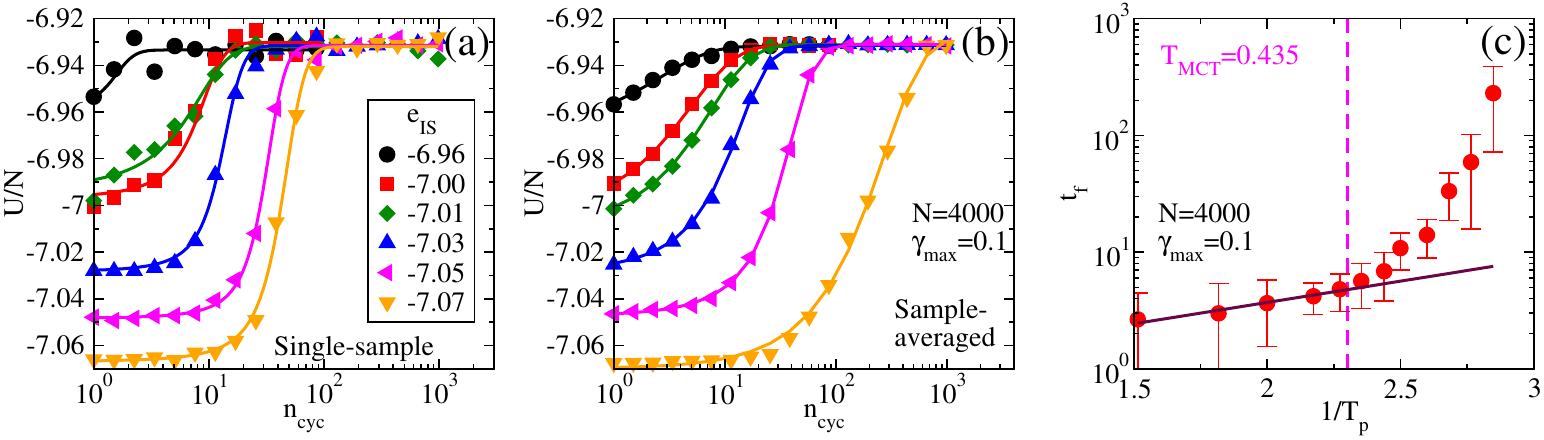}
    \caption{\textbf{Dependence of failure time on annealing:} Potential energy $U/N$ {\it vs.} cycle number $n_{cyc}$ for $N=4000$ and $\gamma_{max} = 0.1$ for different degrees of annealing, expressed in terms of initial inherent structure energy $e_{IS}$ in the legend for (a) single sample and (b) sample averaged data. Solid lines are fits with the sigmoid function. [(a) shares color code with (b)]. (c) Sample-averaged failure time $\langle t_f\rangle $ {\it vs.} degree of annealing, expressed in terms of the parent temperature $T_p$. The error bar represents the sample-to-sample standard deviation. Below the mode coupling temperature $T_{MCT}=0.435$, $t_f$ increases in a super-Arrhenius fashion with decreasing $T_p$.}
    \label{fig:e_vs_ncyc_diffanneal}
\end{figure*}

\noindent{\bf {\em Identification of failure events:}}
For $\gamma_{max} > \gamma_{max}^Y$, the sheared glasses make a transition to a diffusive steady state after 
a finite number of strain cycles, which we refer to as the failure time $t_f$ (we use the number of cycles, $n_{cyc}$, as the time variable, and refer to it interchangeably as time, t).  To identify the failure time, we monitor the per-particle potential energy $U/N$, structural change through the bond-orientational order parameter $q_6$\cite{SteinhardtPRB1983}, and plasticity using the per particle non-affine displacement $D^2_{min}$\cite{FalkPRE1998} from the previous cycle to the present. Details of the calculation of these metrics are given in the SI (Sec: \ref{SI_tf}). 
As shown in Fig. \ref{fig:e_vs_ncyc_diffanneal}(a)-(c) for a well-annealed sample, all the three quantities change sharply when the system fails. The failure time $t_f$ is estimated to be the mid point of the transformation, using a fit to a sigmoid function (SI Sec: \ref{SI_tf}). We define the failure initiation time $t_{fi}=t_f-w/2$ where the width $w$ is the time interval over which the transformation happens. Further details regarding the fitting procedure, and illustrative results for  several  samples across different strain amplitudes, different system sizes, and different degrees of annealing are given in the SI (Sec: \ref{SI_tf}). 


\noindent{\bf {\em Strain amplitude dependence of failure time:}}
We study the dependence of sample-averaged failure time $\langle t_f\rangle$ on the strain amplitude $\gamma_{max}$ obtained from the different metrics. Several forms of the dependence of the fatigue failure times under different loading conditions and settings are discussed in the literature\cite{SureshCUP1998,KunJStat2007,KunPRL2008,Bhowmik_2022}, not all of which correspond to a divergence of the failure time, notwithstanding the possibility of a fatigue limit. We find that the failure time $t_f$ does indeed diverge in a power law fashion as the yield strain amplitude $\gamma_{max}^Y$ is approached from above. As shown in Fig. \ref{fig:failure_time_calc_eIS-7.00_diffmetric}(d), the failure times (obtained from different metrics) can be well described by the diverging power-law behaviour, with a divergence exponent of $-2$: 
\begin{equation}
    \langle t_f \rangle \propto (\gamma_{max}-\gamma_{max}^Y)^{-2}.
\end{equation} 
The robustness of such a power-law is verified across three different system sizes as shown in Fig. \ref{fig:failure_time_calc_eIS-7.00_diffmetric}(e) for well-annealed glasses and in Fig. \ref{fig:failure_time_calc_eIS-7.00_diffmetric}(f) for poorly annealed glasses. The apparent power divergence of cycles to reach steady states has previously been observed in simulations \cite{FioccoPRE2013,Regev2013,LeishanthemNatCom2017,KurotaniCommunMat2022,PRIEZJEV2023112230}, as well as theoretical estimates \cite{ParleyPRL2022}, but with varying exponent estimates \cite{Regev2013,KurotaniCommunMat2022,PRIEZJEV2023112230}. The estimated exponent value reported here is more reliable, we believe, given the significantly better sampling, as well as the robustness with respect to system sizes and degree of annealing. We further confirm these results by estimating the mean failure times from sample to sample fluctuations of energy \cite{CabrioluSM19}, as shown in the SI (Sec. \ref{SI_fluct}).


\begin{figure*}[ht!]
        \centering
        \includegraphics[width=.98\linewidth]{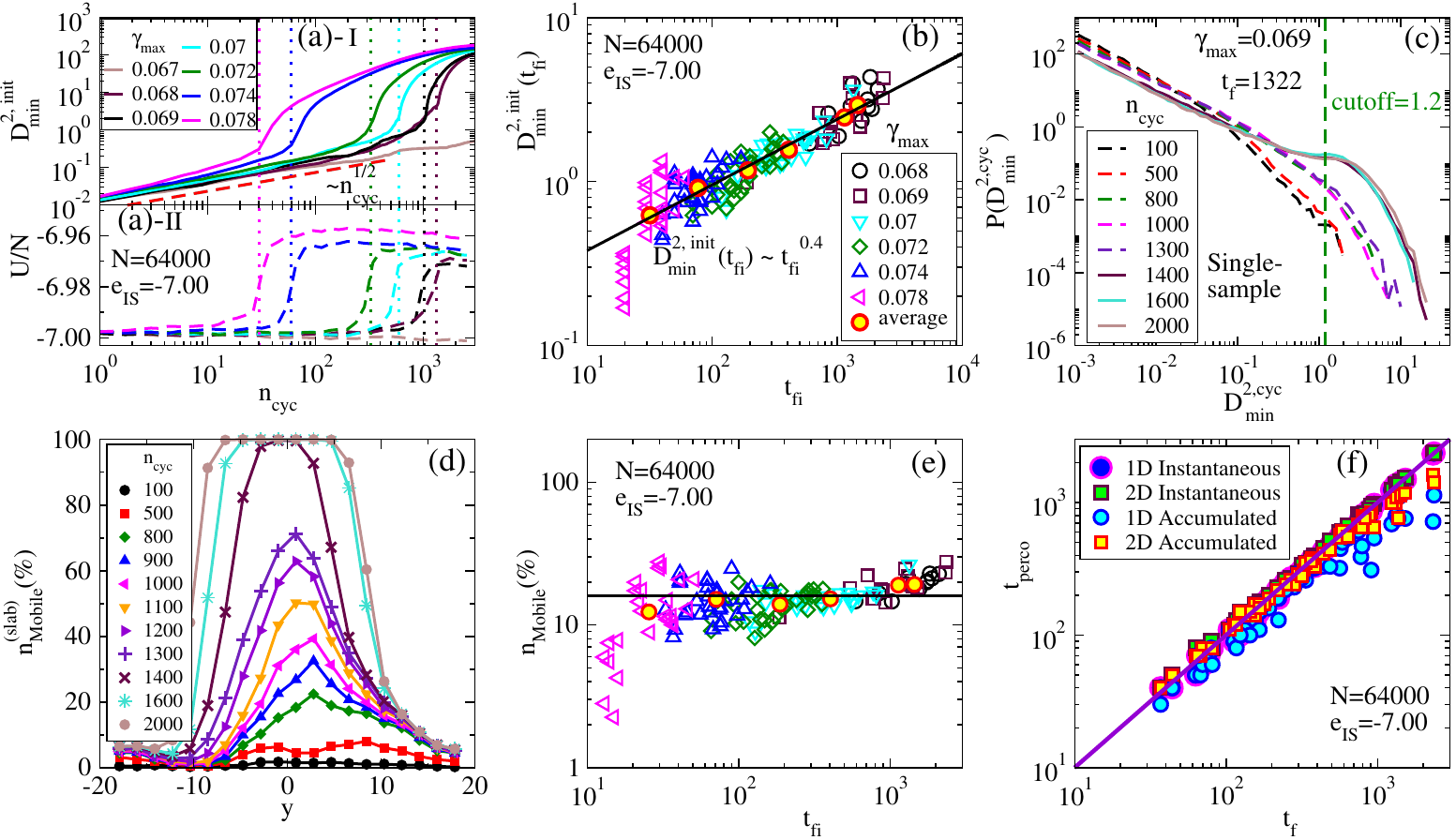}
    \caption{\textbf{Plastic rearrangements and failure times:} Non-affine displacements are used to quantify plasticity and correlations with failure times. The non-affine displacement with respect to the initial configuration $D^{2,init}_{min}$, averaged over particles for a single sample, is shown in (a)-I (top), and the corresponding variation of the energy per particle  $U/N$ is shown in (a)-II (bottom), as a function of the number of cycles $n_{cyc}$ for several $\gamma_{max}$. The failure times are indicated with vertical dotted lines across which both the quantities increase sharply. (b) Accumulated damage till the failure initiation, $D^{2,init}_{min}(t_{fi})$, as a function of the failure initiation time $t_{fi}$ for different $\gamma_{max}$. The solid line is fit to a power law with exponent $0.4$. (c) Distribution of cycle to cycle non-affine displacement $D^{2,cyc}_{min} \equiv D^{2}_{min} (\Delta t = 10\text{ cycle})$, computed at different numbers of cycles of strain. A threshold of $D^{2,cyc}_{min}=1.2$ is chosen beyond which the particles are identified as mobile particles. (d) Accumulation of mobile particles (for $\gamma_{max} = 0.069$) along the gradient (y) direction, showing that plasticity accumulates in a spatially heterogeneous way well before failure, and is concentrated around the eventual location of the shear band that forms at failure. (e) The total number of mobile particles accumulated up to the failure initiation time $t_{fi}$ is plotted against $t_{fi}$ for runs starting with different configurations and different $\gamma_{max}$ [(e) shares color code with (b)]. The fraction of mobile particles reaches a nearly constant value at the failure, for $t_{fi}$ that vary over more than two orders of magnitude. (f) The cycle at which instantaneous or accumulated mobile particles percolate in one or two directions of the system, compared with failure time ($t_f$), showing that the percolation occurs close to the time of failure for instantaneous mobile particles and it occurs before $t_f$ for accumulated mobile particles. Data points for each case are collected over different samples across several $\gamma_{max}$. All the data presented here are for $N=64000$ and $e_{IS}=-7.00$.}
    \label{fig:nActive_usingD2cyc_accumulated}
\end{figure*}

\medskip

\noindent{\bf {\em Dependence of failure time on the degree of annealing:}} The nature of the yielding transition strongly depends on the degree of annealing of the glass \cite{LeishanthemNatCom2017,BhaumikPNAS2021,YehPRL2020,BhaumikJCP2022,ParleyPRL2022}, which prompts us to investigate how the degree of annealing may influence the failure times. Even though the power law exponent for the divergence of failure times is the same for the two cases of annealing described above, the failure time at a given strain amplitude increases with increasing annealing. We thus consider glasses obtained from a wide range of parent temperatures $T_{p}$ (see Methods) that straddles the $T_{p}$ corresponding to the threshold degree of annealing previously identified\cite{BhaumikPNAS2021}, the mode coupling temparature $T_{MCT}$. For a fixed strain amplitude $\gamma_{max}=0.1$, the cycle dependent evolution energy for a single sample and sample-averaged energy  for different degrees of annealing is shown in Fig. \ref{fig:e_vs_ncyc_diffanneal}(a) and (b), respectively. The failure times $t_f$ extracted as describe earlier exhibit Arrhenius dependence on $T_{p}$ for moderate annealing (higher $T_{p}$), similar to observations in \cite{CochranPRL2024} for an elastoplastic model, as shown in Fig. \ref{fig:e_vs_ncyc_diffanneal}(c). However, the dependence becomes super-Arrhenius at low $T_p$. Interestingly, the crossover occurs around $T_p = T_{MCT}$. The significance of this crossover degree of annealing (speculated upon in \cite{BhaumikPNAS2021}) need to be further investigated. 

We also compute the distribution of failure times, whose analysis will be pursued elsewhere. Finally, we also investigate the width (in cycles) of the failure event, and present the results in the SI (Sec. \ref{SI_width}).





\begin{figure*}[!htb]
        \centering
        \includegraphics[width=0.98\linewidth]{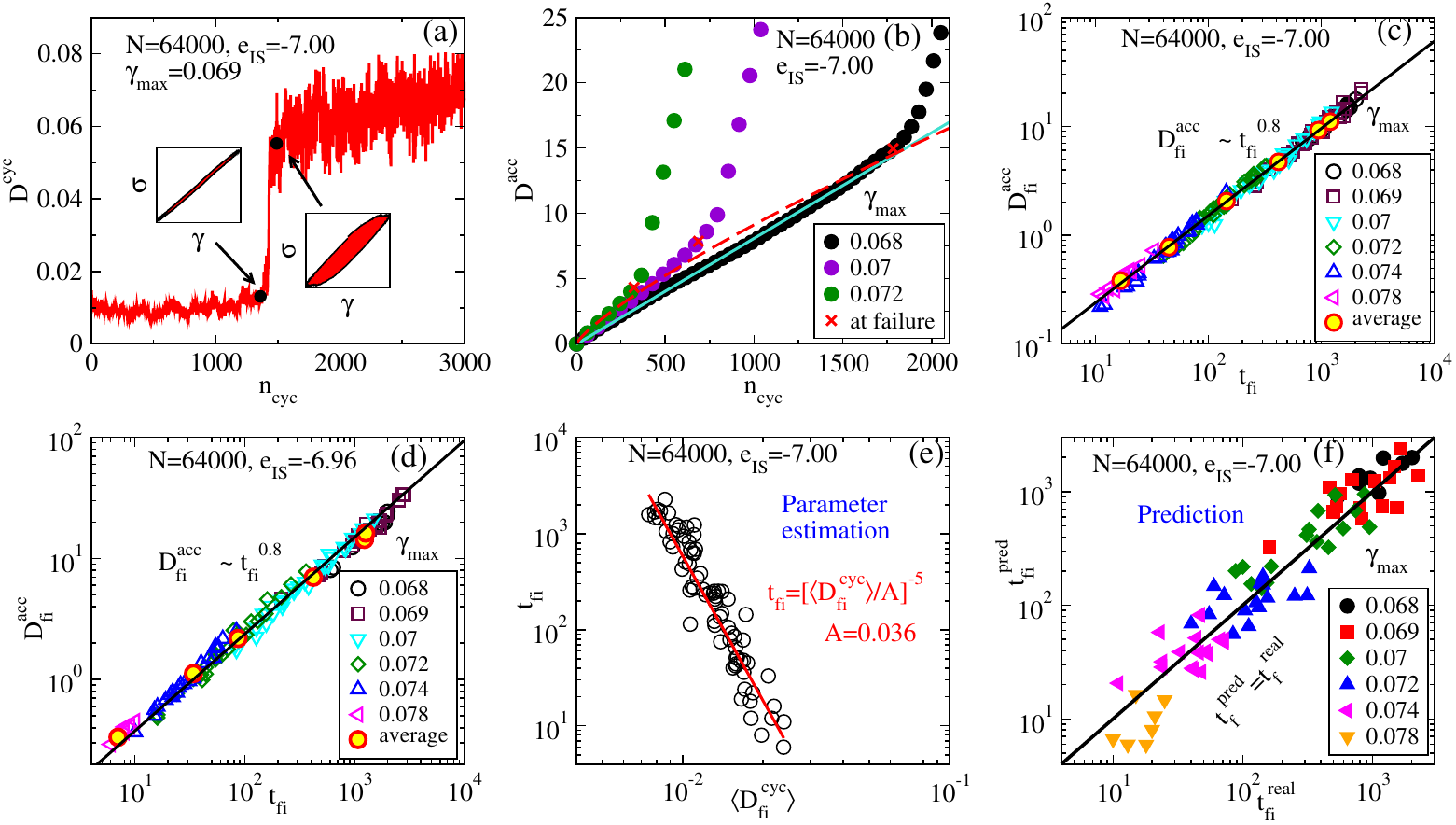}
    \caption{\textbf{Damage, quantified by dissipated energy:} (a) Damage in each cycle $D^{cyc}$, defined as the stress-strain loop area, as a function of strain cycle $n_{cyc}$. The insets show the stress-strain loop at two different cycles across the failure. (b) Accumulated damage $D^{acc}$ against $n_{cyc}$ for different $\gamma_{max}$. The solid line for a representative $\gamma_{max}$ indicates that $D^{acc}$ increases roughly linearly with $n_{cyc}$ till the failure initiation time $t_{fi}$. The crosses indicate failure initiation time $t_{fi}$. The dashed line shows  $D^{acc}_{fi} = A \sim t_{fi}^{0.8}$. Accumulated damage till the failure initiation time $D^{acc}_{fi}$ grows with $t_{fi}$ with a power of $0.8$ for two different degree of annealing (c) $e_{IS}=-7.00$ and (d) $e_{IS}=-6.96$. Symbols with the same color indicate sample-to-sample data for a given $\gamma_{max}$. (e) The failure initiation time $t_{fi}$ {\it vs.} average damage per cycle $\langle D_f^{cyc}\rangle$ (computed up to $t_{fi}$) is fitted as a power law for a subset (50\%) of samples to estimate the parameter $A= 0.036$ for $e_{IS}=-7.00$. 
    (f) The predicted failure initiation time $t_{fi}^{pred}$ based on the the estimated value of parameter $A$ compared with the observed times $t_{fi}^{real}$ (for the remaining $50$\% samples) for different $\gamma_{max}$. 
    }    \label{fig:damage_accumulated_till_failure}
\end{figure*}

\noindent{\bf {\em Mechanism of failure:}} Having characterized the behaviour of failure times, we next focus on possible ways to comprehend the systematic variation as well as the variability of failure times. A notion that has often been discussed in this context is that of damage accumulation \cite{SureshCUP1998,SethnaAnnuRevMaterRes2017,Bhowmik_2022,BhowmikPRE2022,KunJStat2007,KunPRL2008,NaderiProcRSocA2010,JMKrishnan_2023}, that is meant to encapsulate irriversible changes in the solid that eventually lead to failure. 
However, what property can actually be termed as damage for an amorphous system, and how such accumulated damage may account for failure and failure times, is not entirely clear, although the extent of plastic activity is an obvious quantity to consider, and several works ({\it e. g.}, \cite{Bhowmik_2022,NaderiProcRSocA2010}) have considered dissipated work, as quantifiers of damage. We consider now the correlation of failure times and quantifications of plasticity and dissipated work, computing them  up to the failure initiation time $t_{fi}=t_f-w/2$. Accumulating quantities up to $t_{f}$ rather than $t_{fi}$ (the latter being a less ambiguous choice) affect some details of the results, but, the overall picture remains unchanged, as shown in the SI (Sec. \ref{SI_damageupto_tf}).

\noindent{\bf {\em Plastic rearrangements:}} Non-affine displacements have been extensively used in the context of studying plastic rearrangements in glasses. We first compute the non-affine displacement between the initial configuration and the configuration after $n_{cyc}$ cycles of shear, $D^{2,init}_{min}$. The evolution of $D^{2,init}_{min}$ with $n_{cyc}$ is shown in Fig. \ref{fig:nActive_usingD2cyc_accumulated} (a) (top), along with the energy $U/N$ ( Fig. \ref{fig:nActive_usingD2cyc_accumulated} (a) (bottom)), both of which show a rapid rise  across the failure time $t_f$. While enhanced plasticity upon failure is not surprising, we wish to know if the accumulated plastic activity leading up to failure bears any correlation with failure times. To this end, we plot, in Fig. \ref{fig:nActive_usingD2cyc_accumulated} (b) the accumulated plasticity till the failure initiation time, quantified by $D^{2,init}_{min}(t_{fi})$ as a function of the failure time $t_f$ for a range of strain amplitudes. Remarkably, data for all the individual cases broadly fall on the same master curve, that is well described by a power law with exponent $0.4$. Results for poorly annealed glasses are very similar, as shown in the SI (Sec. \ref{SI_d2mininitpoorly}).

Although the accumulated damage as measure by $D^{2,init}_{min}$ does not reach a constant value at failure, as previously argued in other cases \cite{Bhowmik_2022,NaderiProcRSocA2010}, the observed power law relationship across several different samples and strain amplitudes indicates a close relationship. Since $D^{2,init}_{min}$ contains contributions from plastic activity (however, potentially occurring several times in the same regions in space before failure), as well as non-affine displacements that may arise away from the centres of plastic activity, we consider a different measure that seeks to quantify the fraction of a sample in which plastic activity has taken place. To this end, we calculate damage as the accumulation of {\it mobile} particles that undergo plastic rearrangements. Mobile particles are identified through cycle-to-cycle non-affine displacements. In practice we compute $D^{2,cyc}_{min}$ between two stroboscopic configurations that are $10$ cycles apart. The distribution of $D^{2,cyc}_{min}$ for a given amplitude $\gamma_{max} = 0.069$, for different number of strain cycles $n_{cyc}$ are shown in Fig. \ref{fig:nActive_usingD2cyc_accumulated}(c) for a sample with $t_f \approx 1322$. The distributions exhibit power-law behaviour, followed by a cutoff, when $n_{cyc}$ is small. For large $n_{cyc}$, beyond the failure time $t_f \approx 1322$, substantial deviations are apparent beyond $D^{2,cyc}_{min} \approx 1.0$, stemming from very high plastic activity after failure. As in past work ({\it e.g.} \cite{LeishanthemNatCom2017,BhaumikPRL2022} it is necessary to employ a cutoff in order to identify mobile particles, in the absence, however, of a clear bimodality. We consider cutoffs in the range $0.5$ to $2.0$, which provide consistent but numerically different results depending on the choice. There is a need for developing better methods to detect and quantify, let alone predict\cite{Richard_PhysRevMaterials.4.113609}, plasticity. Recent work investigating localized modes and topological characterisation are promising directions, but which need further elaboration  \cite{Lerner2021,WuNatCom2023,Falk_2024_PhysRevE.109.L053002,Zaccone_PNASnexus24}. The results are shown in Fig. \ref{fig:nActive_usingD2cyc_accumulated} (d)-(f) for the cutoff value of  $D^{2,cyc}_{min} = 1.2$. Particles with $D^{2,cyc}_{min}$ greater than the cutoff are labeled as mobile particles, and we consider the cumulative set of mobile particles as a function of cycles. To understand the manner in which plastic activity may cumulatively lead to failure, we consider the fraction of particles that are mobile, in slabs indexed by their position $y$ along the gradient direction. Fig. \ref{fig:nActive_usingD2cyc_accumulated} (d) shows the fraction of mobile particles $n^{(slab)}_{Mobile}(\%)$ {\it vs.} $y$, for different cycles. Although the cycle-to-cycle activity shows no clear spatial pattern before failure, strikingly, we see that there is a heterogeneous accumulation of plastic activity around the eventual position of shear band formation upon failure (shifted to be centered at $y = 0$ for visualization), well before failure occurs. We consider next the fraction of particles in the entire system, $n_{Mobile}(\%)$, up to the failure initiation time $t_{fi}$. This fraction is plotted against $t_{fi}$ in Fig. \ref{fig:nActive_usingD2cyc_accumulated} (e), for several strain amplitudes and samples. Interestingly,  $n_{Mobile}(\%) (t_{fi})$ is independent of the failure time -- the glasses fail when a (nearly) fixed fraction of particles undergo plastic rearrangements! 

There have been several works that have considered the possible relation between yielding or failure and the percolation of regions of plastic activity \cite{Chikkadi_2015,Shrivastav2016,GhoshPRL2017,LeishanthemNatCom2017,SopuPRL2017} but the results so far could be characterised as tentative. Largely, these investigations have been for the case of uniform shear and have considered the percolation of particles that are mobile upon yielding/failure. For cyclically sheared glasses, it was shown in \cite{LeishanthemNatCom2017} that particles that are mobile above yielding percolate, but do so discontinuously. In the present context, what is of interest is to know if the percolation of accumulated plasticity may act as a precursor to failure. We thus consider the accumulated mobile particles, and perform a percolation analysis (as described in \cite{LeishanthemNatCom2017}) after different numbers of cycles. In \ref{fig:nActive_usingD2cyc_accumulated} (f), the time at which the percolation of accumulated mobile particles occurs is shown as a function of the failure time, alongside the time at which percolation occurs for particles that are mobile at a given cycle. The data clearly shows that for all the samples and strain amplitudes studied, the percolation of accumulated mobile particles closely tracks failure, and precedes the percolation of the current subset of mobile particles. The latter observation clearly distinguishes the percolation observed previously \cite{LeishanthemNatCom2017} {\it upon} failure, to the percolation of activity that serves as a precursor to failure. Corresponding results for poorly annealed glasses, shown in the SI (Sec. \ref{SI_d2mininitpoorly}), indicate a poorer correlation, emphasizing the need for improved methods to identify regions of plasticity.


\noindent{\bf {\em Dissipated work:}} As another measure of accumulated damage, we consider dissipated work, that have been studied in several previous works. We define damage in each cycle ($D^{cyc}$) in terms of dissipated work in a mechanical system, i.e. the stress-strain loop area. 
The variation of $D^{cyc}$ with strain cycle is shown in Fig. \ref{fig:damage_accumulated_till_failure}(a). 
Prior to failure, $D^{cyc}$ is small and nearly constant as the stress-strain curve encloses a small area, as shown in the inset of Fig. \ref{fig:damage_accumulated_till_failure}(a). After a sufficient number of strain cycles, when the system fails, $D^{cyc}$ increases suddenly owing to the enhanced plasticity in the yielded system, and the stress-strain curve encloses a large loop area. In Fig. \ref{fig:damage_accumulated_till_failure}(b), we show the evolution of accumulated damage till a cycle $t$, defined as $D^{acc}(t) = \sum_{t'=1}^{t}D^{cyc}(t')$. For each sample, we collect the accumulated damage till the failure initiation time $D^{acc}_{fi}=D^{acc}(t=t_{fi})$, which are shown for well-annealed and poorly-annealed systems, in Fig. \ref{fig:damage_accumulated_till_failure}(c) and (d), respectively.
We find $D^{acc}_{fi}$ increases with $t_{fi}$ following a power-law scaling behavior:
\begin{align}\label{eq:failure_line}
    D^{acc}_{fi} \sim t_{fi}^{\beta}.
\end{align}
where $\beta=0.8$ and is found to be the same for well annealed and poorly annealed glasses. 
Importantly, the exponent being non-zero reveals a positive correlation between $D^{acc}_{fi}$ and $t_{fi}$, and there exists no \textit{critical damage} that the system needs to achieve before it fails. 

Nevertheless, the striking power law relationship we observe in Fig. \ref{fig:damage_accumulated_till_failure}(c) and (d), between $D^{acc}_{fi}$ and $t_{fi}$, together with the nearly linear increase of $D^{acc}$ observed in  Fig. \ref{fig:damage_accumulated_till_failure}(b), permits the prediction of failure times from the behaviour of the damage in the initial cycles, as we now demonstrate. (In passing, we note that whereas  $D^{acc} (t) \sim t$, $D^{2,init}_{min} (t) \sim t^{1/2}$ before failure, and thus, we expect $D^{2,init}_{min}(t_{fi}) \sim D^{acc} (t_{fi})^{1/2} \sim t_{fi}^{\beta/2}$, which we indeed observe.) Firstly, Fig. \ref{fig:damage_accumulated_till_failure}(c) and (d) permit the evaluation of (the annealing dependent) pre-factors in $D^{acc}_{fi} = A t_{fi}^{\beta}$. Knowing the pre-factor $A$, the failure time can be predicted from the cycle-dependent value of $D^{acc}$. Assuming that the damage per cycle, $\langle D^{cyc}\rangle$ is nearly constant till failure, we have $\langle D^{cyc}\rangle = A t_{fi}^{\beta - 1}$ to predict $t_{fi}$ from the observed $\langle D^{cyc}\rangle$ in the initial cycles. To verify this, first, we take half of the samples from our whole ensemble that fail at different time across different $\gamma_{max}$. We compute the slope of the accumulated damage against $n_{cyc}$ up to the failure initiation time $t_{fi}$ of these training samples. As shown in Fig. \ref{fig:damage_accumulated_till_failure}(e), we find $t_{fi}$ decreases with the rate of accumulated damage $ D^{cyc}_{fi}$ (computed up to $t_{fi}$) as 
\begin{align}\label{eq:training}
t_{fi}\simeq \left( \frac{D^{cyc}_{fi}}{A} \right)^ {-\beta^\prime}
\end{align}
with $A \approx 0.036$ and $\beta^{\prime}=5$. Note that the value of $\beta^{\prime}=1/(1-\beta)$ is in excellent agreement with independent measure of $\beta$ from the analysis of $D^{acc}_{fi}$.
Knowing the pre-factor from training samples, we can now test the other half of the samples
to predict $t_{fi}$. For each test sample, we compute the rate of accumulated damage $D^{cyc}_{n}$ for the first $n$-number of cycles and predict the failure initiation time from Eq. \ref{eq:training}. Data presented in Fig. \ref{fig:damage_accumulated_till_failure}(f) for $n=20$ shows the predicted failure initiation times are in excellent agreement with the actual values (Results for the poorly annealed case are shown in SI (Sec. \ref{SI_d2mininitpoorly}). 

Strikingly, our analysis demonstrates that the failure time of a sample can be predicted using the knowledge of only the first few cycles of strain, an observation which can be useful in predicting the failure time of actual materials.

\noindent{\bf {\em Conclusions:}}
In summary, we have studied the fatigue failure phenomenon in a model glass subject to finite rate cyclic shear. Failure is identified through a sigmoidal, rapid, change in energy, measures of structural change, and plasticity. The number of cycles to failure, calculated from these metrics, shows a power-law dependence on the distance from the yield strain amplitude from above,  with a power-law exponent of $-2$. The exponent is also robust across different degrees of annealing and different system sizes. For a constant strain amplitude, the failure time increases as we increase the degree of annealing, with an initial Arrhenius dependence on the parent temperature, followed by a super-Arrhenius dependence at low parent temperatures. The crossover corresponds to the threshold degree of annealing identified earlier, across which the nature of yielding changes qualitatively. Finally, we investigate whether damage accumulated over cycles bears a relationship with failure. Using two metrics based on plasticity and dissipated energy, we find striking relationships between accumulated damage and failure times. We find that the failure occurs when the accumulated subset of (mobile) particles that have undergone plastic rearrangement reaches a fixed fraction, for which we observe a percolation of such accumulated particles. Considering dissipated energy as the measure of damage, although we find no critical level of damage at which failure occurs, we find a robust power law relationship between dissipated energy till failure and the failure time. Such a relationship permits a prediction of failure times based on the response of the solid to initial cycles of shear. Together, these results present an appealing picture of how accumulated damage leads to fatigue failure. These results should also contribute significantly in developing a full-fledged microscopic picture and theoretical explanation of the fatigue failure phenomenon in amorphous solids.

\bibliography{Bibliography}

\begin{thebibliography}{50}%
\makeatletter
\providecommand \@ifxundefined [1]{%
 \@ifx{#1\undefined}
}%
\providecommand \@ifnum [1]{%
 \ifnum #1\expandafter \@firstoftwo
 \else \expandafter \@secondoftwo
 \fi
}%
\providecommand \@ifx [1]{%
 \ifx #1\expandafter \@firstoftwo
 \else \expandafter \@secondoftwo
 \fi
}%
\providecommand \natexlab [1]{#1}%
\providecommand \enquote  [1]{``#1''}%
\providecommand \bibnamefont  [1]{#1}%
\providecommand \bibfnamefont [1]{#1}%
\providecommand \citenamefont [1]{#1}%
\providecommand \href@noop [0]{\@secondoftwo}%
\providecommand \href [0]{\begingroup \@sanitize@url \@href}%
\providecommand \@href[1]{\@@startlink{#1}\@@href}%
\providecommand \@@href[1]{\endgroup#1\@@endlink}%
\providecommand \@sanitize@url [0]{\catcode `\\12\catcode `\$12\catcode
  `\&12\catcode `\#12\catcode `\^12\catcode `\_12\catcode `\%12\relax}%
\providecommand \@@startlink[1]{}%
\providecommand \@@endlink[0]{}%
\providecommand \url  [0]{\begingroup\@sanitize@url \@url }%
\providecommand \@url [1]{\endgroup\@href {#1}{\urlprefix }}%
\providecommand \urlprefix  [0]{URL }%
\providecommand \Eprint [0]{\href }%
\providecommand \doibase [0]{https://doi.org/}%
\providecommand \selectlanguage [0]{\@gobble}%
\providecommand \bibinfo  [0]{\@secondoftwo}%
\providecommand \bibfield  [0]{\@secondoftwo}%
\providecommand \translation [1]{[#1]}%
\providecommand \BibitemOpen [0]{}%
\providecommand \bibitemStop [0]{}%
\providecommand \bibitemNoStop [0]{.\EOS\space}%
\providecommand \EOS [0]{\spacefactor3000\relax}%
\providecommand \BibitemShut  [1]{\csname bibitem#1\endcsname}%
\let\auto@bib@innerbib\@empty
\bibitem [{\citenamefont {Suresh}(1998)}]{SureshCUP1998}%
  \BibitemOpen
  \bibfield  {author} {\bibinfo {author} {\bibfnamefont {S.}~\bibnamefont
  {Suresh}},\ }\href {https://doi.org/10.1017/CBO9780511806575} {\emph
  {\bibinfo {title} {Fatigue of Materials}}},\ \bibinfo {edition} {2nd}\ ed.\
  (\bibinfo  {publisher} {Cambridge University Press},\ \bibinfo {year}
  {1998})\BibitemShut {NoStop}%
\bibitem [{\citenamefont {Christensen}(2013)}]{Christensen2014}%
  \BibitemOpen
  \bibfield  {author} {\bibinfo {author} {\bibfnamefont {R.~M.}\ \bibnamefont
  {Christensen}},\ }\href@noop {} {\emph {\bibinfo {title} {The Theory of
  Materials Failure}}},\ \bibinfo {edition} {1st}\ ed.\ (\bibinfo  {publisher}
  {Oxford University Press},\ \bibinfo {year} {2013})\BibitemShut {NoStop}%
\bibitem [{\citenamefont {Naderi}\ \emph {et~al.}(2010)\citenamefont {Naderi},
  \citenamefont {Amiri},\ and\ \citenamefont {Khonsari}}]{NaderiProcRSocA2010}%
  \BibitemOpen
  \bibfield  {author} {\bibinfo {author} {\bibfnamefont {M.}~\bibnamefont
  {Naderi}}, \bibinfo {author} {\bibfnamefont {M.}~\bibnamefont {Amiri}},\ and\
  \bibinfo {author} {\bibfnamefont {M.~M.}\ \bibnamefont {Khonsari}},\
  }\bibfield  {title} {\bibinfo {title} {On the thermodynamic entropy of
  fatigue fracture},\ }\href {https://doi.org/10.1098/rspa.2009.0348}
  {\bibfield  {journal} {\bibinfo  {journal} {Proceedings of the Royal Society
  A: Mathematical, Physical and Engineering Sciences}\ }\textbf {\bibinfo
  {volume} {466}},\ \bibinfo {pages} {423} (\bibinfo {year}
  {2010})}\BibitemShut {NoStop}%
\bibitem [{\citenamefont {Mangalath~Shine}\ \emph {et~al.}(2023)\citenamefont
  {Mangalath~Shine}, \citenamefont {Sanchana}, \citenamefont {Padmarekha},
  \citenamefont {Leischner}, \citenamefont {Wellner},\ and\ \citenamefont
  {Murali~Krishnan}}]{JMKrishnan_2023}%
  \BibitemOpen
  \bibfield  {author} {\bibinfo {author} {\bibfnamefont {A.}~\bibnamefont
  {Mangalath~Shine}}, \bibinfo {author} {\bibfnamefont {I.~C.}\ \bibnamefont
  {Sanchana}}, \bibinfo {author} {\bibfnamefont {A.}~\bibnamefont
  {Padmarekha}}, \bibinfo {author} {\bibfnamefont {S.}~\bibnamefont
  {Leischner}}, \bibinfo {author} {\bibfnamefont {F.}~\bibnamefont {Wellner}},\
  and\ \bibinfo {author} {\bibfnamefont {J.}~\bibnamefont {Murali~Krishnan}},\
  }\bibfield  {title} {\bibinfo {title} {Quantification of viscous and damage
  dissipation of bituminous binder and mastic using white-metzner model},\
  }\href {https://doi.org/10.1080/10298436.2023.2238112} {\bibfield  {journal}
  {\bibinfo  {journal} {International Journal of Pavement Engineering}\
  }\textbf {\bibinfo {volume} {24}},\ \bibinfo {pages} {2238112} (\bibinfo
  {year} {2023})}\BibitemShut {NoStop}%
\bibitem [{\citenamefont {Bhowmik}\ \emph
  {et~al.}(2022{\natexlab{a}})\citenamefont {Bhowmik}, \citenamefont
  {Hentchel},\ and\ \citenamefont {Procaccia}}]{Bhowmik_2022}%
  \BibitemOpen
  \bibfield  {author} {\bibinfo {author} {\bibfnamefont {B.~P.}\ \bibnamefont
  {Bhowmik}}, \bibinfo {author} {\bibfnamefont {H.~G.~E.}\ \bibnamefont
  {Hentchel}},\ and\ \bibinfo {author} {\bibfnamefont {I.}~\bibnamefont
  {Procaccia}},\ }\bibfield  {title} {\bibinfo {title} {Fatigue and collapse of
  cyclically bent strip of amorphous solid},\ }\href
  {https://doi.org/10.1209/0295-5075/ac4ba5} {\bibfield  {journal} {\bibinfo
  {journal} {Europhysics Letters}\ }\textbf {\bibinfo {volume} {137}},\
  \bibinfo {pages} {46002} (\bibinfo {year} {2022}{\natexlab{a}})}\BibitemShut
  {NoStop}%
\bibitem [{\citenamefont {Bhowmik}\ \emph
  {et~al.}(2022{\natexlab{b}})\citenamefont {Bhowmik}, \citenamefont
  {Hentschel},\ and\ \citenamefont {Procaccia}}]{BhowmikPRE2022}%
  \BibitemOpen
  \bibfield  {author} {\bibinfo {author} {\bibfnamefont {B.~P.}\ \bibnamefont
  {Bhowmik}}, \bibinfo {author} {\bibfnamefont {H.~G.~E.}\ \bibnamefont
  {Hentschel}},\ and\ \bibinfo {author} {\bibfnamefont {I.}~\bibnamefont
  {Procaccia}},\ }\bibfield  {title} {\bibinfo {title} {Scaling theory for
  w\"ohler plots in amorphous solids under cyclic forcing},\ }\href
  {https://doi.org/10.1103/PhysRevE.105.015001} {\bibfield  {journal} {\bibinfo
   {journal} {Phys. Rev. E}\ }\textbf {\bibinfo {volume} {105}},\ \bibinfo
  {pages} {015001} (\bibinfo {year} {2022}{\natexlab{b}})}\BibitemShut
  {NoStop}%
\bibitem [{\citenamefont {Kun}\ \emph {et~al.}(2007)\citenamefont {Kun},
  \citenamefont {Costa}, \citenamefont {Filho}, \citenamefont {Andrade},
  \citenamefont {Soares}, \citenamefont {Zapperi},\ and\ \citenamefont
  {Herrmann}}]{KunJStat2007}%
  \BibitemOpen
  \bibfield  {author} {\bibinfo {author} {\bibfnamefont {F.}~\bibnamefont
  {Kun}}, \bibinfo {author} {\bibfnamefont {M.~H.}\ \bibnamefont {Costa}},
  \bibinfo {author} {\bibfnamefont {R.~N.~C.}\ \bibnamefont {Filho}}, \bibinfo
  {author} {\bibfnamefont {J.~S.}\ \bibnamefont {Andrade}}, \bibinfo {author}
  {\bibfnamefont {J.~B.}\ \bibnamefont {Soares}}, \bibinfo {author}
  {\bibfnamefont {S.}~\bibnamefont {Zapperi}},\ and\ \bibinfo {author}
  {\bibfnamefont {H.~J.}\ \bibnamefont {Herrmann}},\ }\bibfield  {title}
  {\bibinfo {title} {Fatigue failure of disordered materials},\ }\href
  {https://doi.org/10.1088/1742-5468/2007/02/P02003} {\bibfield  {journal}
  {\bibinfo  {journal} {Journal of Statistical Mechanics: Theory and
  Experiment}\ }\textbf {\bibinfo {volume} {2007}},\ \bibinfo {pages} {P02003}
  (\bibinfo {year} {2007})}\BibitemShut {NoStop}%
\bibitem [{\citenamefont {Leocmach}\ \emph {et~al.}(2014)\citenamefont
  {Leocmach}, \citenamefont {Perge}, \citenamefont {Divoux},\ and\
  \citenamefont {Manneville}}]{LeomachPRL14}%
  \BibitemOpen
  \bibfield  {author} {\bibinfo {author} {\bibfnamefont {M.}~\bibnamefont
  {Leocmach}}, \bibinfo {author} {\bibfnamefont {C.}~\bibnamefont {Perge}},
  \bibinfo {author} {\bibfnamefont {T.}~\bibnamefont {Divoux}},\ and\ \bibinfo
  {author} {\bibfnamefont {S.}~\bibnamefont {Manneville}},\ }\bibfield  {title}
  {\bibinfo {title} {Creep and fracture of a protein gel under stress},\ }\href
  {https://doi.org/10.1103/PhysRevLett.113.038303} {\bibfield  {journal}
  {\bibinfo  {journal} {Phys. Rev. Lett.}\ }\textbf {\bibinfo {volume} {113}},\
  \bibinfo {pages} {038303} (\bibinfo {year} {2014})}\BibitemShut {NoStop}%
\bibitem [{\citenamefont {Aime}\ \emph {et~al.}(2018)\citenamefont {Aime},
  \citenamefont {Ramos},\ and\ \citenamefont
  {Cipelletti}}]{aime2018microscopic}%
  \BibitemOpen
  \bibfield  {author} {\bibinfo {author} {\bibfnamefont {S.}~\bibnamefont
  {Aime}}, \bibinfo {author} {\bibfnamefont {L.}~\bibnamefont {Ramos}},\ and\
  \bibinfo {author} {\bibfnamefont {L.}~\bibnamefont {Cipelletti}},\ }\bibfield
   {title} {\bibinfo {title} {Microscopic dynamics and failure precursors of a
  gel under mechanical load},\ }\href@noop {} {\bibfield  {journal} {\bibinfo
  {journal} {Proceedings of the National Academy of Sciences}\ }\textbf
  {\bibinfo {volume} {115}},\ \bibinfo {pages} {3587} (\bibinfo {year}
  {2018})}\BibitemShut {NoStop}%
\bibitem [{\citenamefont {Cipelletti}\ \emph {et~al.}(2020)\citenamefont
  {Cipelletti}, \citenamefont {Martens},\ and\ \citenamefont
  {Ramos}}]{Cipelletti2020}%
  \BibitemOpen
  \bibfield  {author} {\bibinfo {author} {\bibfnamefont {L.}~\bibnamefont
  {Cipelletti}}, \bibinfo {author} {\bibfnamefont {K.}~\bibnamefont
  {Martens}},\ and\ \bibinfo {author} {\bibfnamefont {L.}~\bibnamefont
  {Ramos}},\ }\bibfield  {title} {\bibinfo {title} {Microscopic precursors of
  failure in soft matter},\ }\href {https://doi.org/10.1039/C9SM01730E}
  {\bibfield  {journal} {\bibinfo  {journal} {Soft Matter}\ }\textbf {\bibinfo
  {volume} {16}},\ \bibinfo {pages} {82} (\bibinfo {year} {2020})}\BibitemShut
  {NoStop}%
\bibitem [{\citenamefont {Bonn}\ \emph {et~al.}(2017)\citenamefont {Bonn},
  \citenamefont {Denn}, \citenamefont {Berthier}, \citenamefont {Divoux},\ and\
  \citenamefont {Manneville}}]{BonnRevModPhys2017}%
  \BibitemOpen
  \bibfield  {author} {\bibinfo {author} {\bibfnamefont {D.}~\bibnamefont
  {Bonn}}, \bibinfo {author} {\bibfnamefont {M.~M.}\ \bibnamefont {Denn}},
  \bibinfo {author} {\bibfnamefont {L.}~\bibnamefont {Berthier}}, \bibinfo
  {author} {\bibfnamefont {T.}~\bibnamefont {Divoux}},\ and\ \bibinfo {author}
  {\bibfnamefont {S.}~\bibnamefont {Manneville}},\ }\bibfield  {title}
  {\bibinfo {title} {Yield stress materials in soft condensed matter},\ }\href
  {https://doi.org/10.1103/RevModPhys.89.035005} {\bibfield  {journal}
  {\bibinfo  {journal} {Rev. Mod. Phys.}\ }\textbf {\bibinfo {volume} {89}},\
  \bibinfo {pages} {035005} (\bibinfo {year} {2017})}\BibitemShut {NoStop}%
\bibitem [{\citenamefont {Nicolas}\ \emph {et~al.}(2018)\citenamefont
  {Nicolas}, \citenamefont {Ferrero}, \citenamefont {Martens},\ and\
  \citenamefont {Barrat}}]{NicolasRevModPhys2018}%
  \BibitemOpen
  \bibfield  {author} {\bibinfo {author} {\bibfnamefont {A.}~\bibnamefont
  {Nicolas}}, \bibinfo {author} {\bibfnamefont {E.~E.}\ \bibnamefont
  {Ferrero}}, \bibinfo {author} {\bibfnamefont {K.}~\bibnamefont {Martens}},\
  and\ \bibinfo {author} {\bibfnamefont {J.-L.}\ \bibnamefont {Barrat}},\
  }\bibfield  {title} {\bibinfo {title} {Deformation and flow of amorphous
  solids: Insights from elastoplastic models},\ }\href
  {https://doi.org/10.1103/RevModPhys.90.045006} {\bibfield  {journal}
  {\bibinfo  {journal} {Rev. Mod. Phys.}\ }\textbf {\bibinfo {volume} {90}},\
  \bibinfo {pages} {045006} (\bibinfo {year} {2018})}\BibitemShut {NoStop}%
\bibitem [{\citenamefont {Kumar}\ and\ \citenamefont
  {Procaccia}(2024)}]{Kumar_2024}%
  \BibitemOpen
  \bibfield  {author} {\bibinfo {author} {\bibfnamefont {A.}~\bibnamefont
  {Kumar}}\ and\ \bibinfo {author} {\bibfnamefont {I.}~\bibnamefont
  {Procaccia}},\ }\bibfield  {title} {\bibinfo {title} {Elasticity, plasticity
  and screening in amorphous solids: A short review},\ }\href
  {https://doi.org/10.1209/0295-5075/ad2087} {\bibfield  {journal} {\bibinfo
  {journal} {Europhysics Letters}\ }\textbf {\bibinfo {volume} {145}},\
  \bibinfo {pages} {26002} (\bibinfo {year} {2024})}\BibitemShut {NoStop}%
\bibitem [{\citenamefont {Sollich}(2024)}]{SollichArxiv2024}%
  \BibitemOpen
  \bibfield  {author} {\bibinfo {author} {\bibfnamefont {P.}~\bibnamefont
  {Sollich}},\ }\href@noop {} {\bibinfo {title} {Challenges in the rheology of
  glasses}} (\bibinfo {year} {2024}),\ \Eprint
  {https://arxiv.org/abs/2401.16409} {arXiv:2401.16409 [cond-mat.soft]}
  \BibitemShut {NoStop}%
\bibitem [{\citenamefont {Lerner}(2019)}]{LernerJNonCrys2019}%
  \BibitemOpen
  \bibfield  {author} {\bibinfo {author} {\bibfnamefont {E.}~\bibnamefont
  {Lerner}},\ }\bibfield  {title} {\bibinfo {title} {Mechanical properties of
  simple computer glasses},\ }\href
  {https://doi.org/https://doi.org/10.1016/j.jnoncrysol.2019.119570} {\bibfield
   {journal} {\bibinfo  {journal} {Journal of Non-Crystalline Solids}\ }\textbf
  {\bibinfo {volume} {522}},\ \bibinfo {pages} {119570} (\bibinfo {year}
  {2019})}\BibitemShut {NoStop}%
\bibitem [{\citenamefont {Fiocco}\ \emph {et~al.}(2013)\citenamefont {Fiocco},
  \citenamefont {Foffi},\ and\ \citenamefont {Sastry}}]{FioccoPRE2013}%
  \BibitemOpen
  \bibfield  {author} {\bibinfo {author} {\bibfnamefont {D.}~\bibnamefont
  {Fiocco}}, \bibinfo {author} {\bibfnamefont {G.}~\bibnamefont {Foffi}},\ and\
  \bibinfo {author} {\bibfnamefont {S.}~\bibnamefont {Sastry}},\ }\bibfield
  {title} {\bibinfo {title} {Oscillatory athermal quasistatic deformation of a
  model glass},\ }\href {https://doi.org/10.1103/PhysRevE.88.020301} {\bibfield
   {journal} {\bibinfo  {journal} {Phys. Rev. E}\ }\textbf {\bibinfo {volume}
  {88}},\ \bibinfo {pages} {020301} (\bibinfo {year} {2013})}\BibitemShut
  {NoStop}%
\bibitem [{\citenamefont {Regev}\ \emph {et~al.}(2013)\citenamefont {Regev},
  \citenamefont {Lookman},\ and\ \citenamefont {Reichhardt}}]{Regev2013}%
  \BibitemOpen
  \bibfield  {author} {\bibinfo {author} {\bibfnamefont {I.}~\bibnamefont
  {Regev}}, \bibinfo {author} {\bibfnamefont {T.}~\bibnamefont {Lookman}},\
  and\ \bibinfo {author} {\bibfnamefont {C.}~\bibnamefont {Reichhardt}},\
  }\bibfield  {title} {\bibinfo {title} {Onset of irreversibility and chaos in
  amorphous solids under periodic shear},\ }\href
  {https://doi.org/10.1103/PhysRevE.88.062401} {\bibfield  {journal} {\bibinfo
  {journal} {Phys. Rev. E}\ }\textbf {\bibinfo {volume} {88}},\ \bibinfo
  {pages} {062401} (\bibinfo {year} {2013})}\BibitemShut {NoStop}%
\bibitem [{\citenamefont {Priezjev}(2013)}]{Priezjev2013}%
  \BibitemOpen
  \bibfield  {author} {\bibinfo {author} {\bibfnamefont {N.~V.}\ \bibnamefont
  {Priezjev}},\ }\bibfield  {title} {\bibinfo {title} {Heterogeneous relaxation
  dynamics in amorphous materials under cyclic loading},\ }\href
  {https://doi.org/10.1103/PhysRevE.87.052302} {\bibfield  {journal} {\bibinfo
  {journal} {Phys. Rev. E}\ }\textbf {\bibinfo {volume} {87}},\ \bibinfo
  {pages} {052302} (\bibinfo {year} {2013})}\BibitemShut {NoStop}%
\bibitem [{\citenamefont {Fiocco}\ \emph {et~al.}(2015)\citenamefont {Fiocco},
  \citenamefont {Foffi},\ and\ \citenamefont {Sastry}}]{FioccoIOP2015}%
  \BibitemOpen
  \bibfield  {author} {\bibinfo {author} {\bibfnamefont {D.}~\bibnamefont
  {Fiocco}}, \bibinfo {author} {\bibfnamefont {G.}~\bibnamefont {Foffi}},\ and\
  \bibinfo {author} {\bibfnamefont {S.}~\bibnamefont {Sastry}},\ }\bibfield
  {title} {\bibinfo {title} {Memory effects in schematic models of glasses
  subjected to oscillatory deformation},\ }\href
  {https://doi.org/10.1088/0953-8984/27/19/194130} {\bibfield  {journal}
  {\bibinfo  {journal} {Journal of Physics: Condensed Matter}\ }\textbf
  {\bibinfo {volume} {27}},\ \bibinfo {pages} {194130} (\bibinfo {year}
  {2015})}\BibitemShut {NoStop}%
\bibitem [{\citenamefont {Leishangthem}\ \emph {et~al.}(2017)\citenamefont
  {Leishangthem}, \citenamefont {Parmar},\ and\ \citenamefont
  {Sastry}}]{LeishanthemNatCom2017}%
  \BibitemOpen
  \bibfield  {author} {\bibinfo {author} {\bibfnamefont {P.}~\bibnamefont
  {Leishangthem}}, \bibinfo {author} {\bibfnamefont {A.}~\bibnamefont
  {Parmar}},\ and\ \bibinfo {author} {\bibfnamefont {S.}~\bibnamefont
  {Sastry}},\ }\bibfield  {title} {\bibinfo {title} {The yielding transition in
  amorphous solids under oscillatory shear deformation},\ }\href
  {https://doi.org/https://doi.org/10.1038} {\bibfield  {journal} {\bibinfo
  {journal} {Nat Commun}\ }\textbf {\bibinfo {volume} {8}},\ \bibinfo {pages}
  {14653} (\bibinfo {year} {2017})}\BibitemShut {NoStop}%
\bibitem [{\citenamefont {Kawasaki}\ and\ \citenamefont
  {Berthier}(2016)}]{kawasakiPRE16}%
  \BibitemOpen
  \bibfield  {author} {\bibinfo {author} {\bibfnamefont {T.}~\bibnamefont
  {Kawasaki}}\ and\ \bibinfo {author} {\bibfnamefont {L.}~\bibnamefont
  {Berthier}},\ }\bibfield  {title} {\bibinfo {title} {Macroscopic yielding in
  jammed solids is accompanied by a nonequilibrium first-order transition in
  particle trajectories},\ }\href {https://doi.org/10.1103/PhysRevE.94.022615}
  {\bibfield  {journal} {\bibinfo  {journal} {Phys. Rev. E}\ }\textbf {\bibinfo
  {volume} {94}},\ \bibinfo {pages} {022615} (\bibinfo {year}
  {2016})}\BibitemShut {NoStop}%
\bibitem [{\citenamefont {Parmar}\ \emph {et~al.}(2019)\citenamefont {Parmar},
  \citenamefont {Kumar},\ and\ \citenamefont {Sastry}}]{ParmarPRX2019}%
  \BibitemOpen
  \bibfield  {author} {\bibinfo {author} {\bibfnamefont {A.~D.~S.}\
  \bibnamefont {Parmar}}, \bibinfo {author} {\bibfnamefont {S.}~\bibnamefont
  {Kumar}},\ and\ \bibinfo {author} {\bibfnamefont {S.}~\bibnamefont
  {Sastry}},\ }\bibfield  {title} {\bibinfo {title} {Strain localization above
  the yielding point in cyclically deformed glasses},\ }\href
  {https://doi.org/10.1103/PhysRevX.9.021018} {\bibfield  {journal} {\bibinfo
  {journal} {Phys. Rev. X}\ }\textbf {\bibinfo {volume} {9}},\ \bibinfo {pages}
  {021018} (\bibinfo {year} {2019})}\BibitemShut {NoStop}%
\bibitem [{\citenamefont {Bhaumik}\ \emph {et~al.}(2021)\citenamefont
  {Bhaumik}, \citenamefont {Foffi},\ and\ \citenamefont
  {Sastry}}]{BhaumikPNAS2021}%
  \BibitemOpen
  \bibfield  {author} {\bibinfo {author} {\bibfnamefont {H.}~\bibnamefont
  {Bhaumik}}, \bibinfo {author} {\bibfnamefont {G.}~\bibnamefont {Foffi}},\
  and\ \bibinfo {author} {\bibfnamefont {S.}~\bibnamefont {Sastry}},\
  }\bibfield  {title} {\bibinfo {title} {The role of annealing in determining
  the yielding behavior of glasses under cyclic shear deformation},\ }\href
  {https://doi.org/https://doi.org/10.1073/pnas.210022711} {\bibfield
  {journal} {\bibinfo  {journal} {PNAS}\ }\textbf {\bibinfo {volume} {118}},\
  \bibinfo {pages} {e2100227118} (\bibinfo {year} {2021})}\BibitemShut
  {NoStop}%
\bibitem [{\citenamefont {Yeh}\ \emph {et~al.}(2020)\citenamefont {Yeh},
  \citenamefont {Ozawa}, \citenamefont {Miyazaki}, \citenamefont {Kawasaki},\
  and\ \citenamefont {Berthier}}]{YehPRL2020}%
  \BibitemOpen
  \bibfield  {author} {\bibinfo {author} {\bibfnamefont {W.-T.}\ \bibnamefont
  {Yeh}}, \bibinfo {author} {\bibfnamefont {M.}~\bibnamefont {Ozawa}}, \bibinfo
  {author} {\bibfnamefont {K.}~\bibnamefont {Miyazaki}}, \bibinfo {author}
  {\bibfnamefont {T.}~\bibnamefont {Kawasaki}},\ and\ \bibinfo {author}
  {\bibfnamefont {L.}~\bibnamefont {Berthier}},\ }\bibfield  {title} {\bibinfo
  {title} {Glass stability changes the nature of yielding under oscillatory
  shear},\ }\href {https://doi.org/10.1103/PhysRevLett.124.225502} {\bibfield
  {journal} {\bibinfo  {journal} {Phys. Rev. Lett.}\ }\textbf {\bibinfo
  {volume} {124}},\ \bibinfo {pages} {225502} (\bibinfo {year}
  {2020})}\BibitemShut {NoStop}%
\bibitem [{\citenamefont {Sastry}(2021)}]{SastryPRL2021}%
  \BibitemOpen
  \bibfield  {author} {\bibinfo {author} {\bibfnamefont {S.}~\bibnamefont
  {Sastry}},\ }\bibfield  {title} {\bibinfo {title} {Models for the yielding
  behavior of amorphous solids},\ }\href
  {https://doi.org/10.1103/PhysRevLett.126.255501} {\bibfield  {journal}
  {\bibinfo  {journal} {Phys. Rev. Lett.}\ }\textbf {\bibinfo {volume} {126}},\
  \bibinfo {pages} {255501} (\bibinfo {year} {2021})}\BibitemShut {NoStop}%
\bibitem [{\citenamefont {Mungan}\ and\ \citenamefont
  {Sastry}(2021)}]{MunganPRL2021}%
  \BibitemOpen
  \bibfield  {author} {\bibinfo {author} {\bibfnamefont {M.}~\bibnamefont
  {Mungan}}\ and\ \bibinfo {author} {\bibfnamefont {S.}~\bibnamefont
  {Sastry}},\ }\bibfield  {title} {\bibinfo {title} {Metastability as a
  mechanism for yielding in amorphous solids under cyclic shear},\ }\href
  {https://doi.org/10.1103/PhysRevLett.127.248002} {\bibfield  {journal}
  {\bibinfo  {journal} {Phys. Rev. Lett.}\ }\textbf {\bibinfo {volume} {127}},\
  \bibinfo {pages} {248002} (\bibinfo {year} {2021})}\BibitemShut {NoStop}%
\bibitem [{\citenamefont {Liu}\ \emph {et~al.}(2022)\citenamefont {Liu},
  \citenamefont {Ferrero}, \citenamefont {Jagla}, \citenamefont {Martens},
  \citenamefont {Rosso},\ and\ \citenamefont {Talon}}]{LiuJCP2022}%
  \BibitemOpen
  \bibfield  {author} {\bibinfo {author} {\bibfnamefont {C.}~\bibnamefont
  {Liu}}, \bibinfo {author} {\bibfnamefont {E.~E.}\ \bibnamefont {Ferrero}},
  \bibinfo {author} {\bibfnamefont {E.~A.}\ \bibnamefont {Jagla}}, \bibinfo
  {author} {\bibfnamefont {K.}~\bibnamefont {Martens}}, \bibinfo {author}
  {\bibfnamefont {A.}~\bibnamefont {Rosso}},\ and\ \bibinfo {author}
  {\bibfnamefont {L.}~\bibnamefont {Talon}},\ }\bibfield  {title} {\bibinfo
  {title} {{The fate of shear-oscillated amorphous solids}},\ }\href
  {https://doi.org/10.1063/5.0079460} {\bibfield  {journal} {\bibinfo
  {journal} {The Journal of Chemical Physics}\ }\textbf {\bibinfo {volume}
  {156}},\ \bibinfo {pages} {104902} (\bibinfo {year} {2022})}\BibitemShut
  {NoStop}%
\bibitem [{\citenamefont {Parley}\ \emph {et~al.}(2022)\citenamefont {Parley},
  \citenamefont {Sastry},\ and\ \citenamefont {Sollich}}]{ParleyPRL2022}%
  \BibitemOpen
  \bibfield  {author} {\bibinfo {author} {\bibfnamefont {J.~T.}\ \bibnamefont
  {Parley}}, \bibinfo {author} {\bibfnamefont {S.}~\bibnamefont {Sastry}},\
  and\ \bibinfo {author} {\bibfnamefont {P.}~\bibnamefont {Sollich}},\
  }\bibfield  {title} {\bibinfo {title} {Mean-field theory of yielding under
  oscillatory shear},\ }\href {https://doi.org/10.1103/PhysRevLett.128.198001}
  {\bibfield  {journal} {\bibinfo  {journal} {Phys. Rev. Lett.}\ }\textbf
  {\bibinfo {volume} {128}},\ \bibinfo {pages} {198001} (\bibinfo {year}
  {2022})}\BibitemShut {NoStop}%
\bibitem [{\citenamefont {Cochran}\ \emph {et~al.}(2024)\citenamefont
  {Cochran}, \citenamefont {Callaghan}, \citenamefont {Caven},\ and\
  \citenamefont {Fielding}}]{CochranPRL2024}%
  \BibitemOpen
  \bibfield  {author} {\bibinfo {author} {\bibfnamefont {J.~O.}\ \bibnamefont
  {Cochran}}, \bibinfo {author} {\bibfnamefont {G.~L.}\ \bibnamefont
  {Callaghan}}, \bibinfo {author} {\bibfnamefont {M.~J.~G.}\ \bibnamefont
  {Caven}},\ and\ \bibinfo {author} {\bibfnamefont {S.~M.}\ \bibnamefont
  {Fielding}},\ }\bibfield  {title} {\bibinfo {title} {Slow fatigue and highly
  delayed yielding via shear banding in oscillatory shear},\ }\href
  {https://doi.org/10.1103/PhysRevLett.132.168202} {\bibfield  {journal}
  {\bibinfo  {journal} {Phys. Rev. Lett.}\ }\textbf {\bibinfo {volume} {132}},\
  \bibinfo {pages} {168202} (\bibinfo {year} {2024})}\BibitemShut {NoStop}%
\bibitem [{\citenamefont {Priezjev}(2023)}]{PRIEZJEV2023112230}%
  \BibitemOpen
  \bibfield  {author} {\bibinfo {author} {\bibfnamefont {N.~V.}\ \bibnamefont
  {Priezjev}},\ }\bibfield  {title} {\bibinfo {title} {Fatigue failure of
  amorphous alloys under cyclic shear deformation},\ }\href
  {https://doi.org/https://doi.org/10.1016/j.commatsci.2023.112230} {\bibfield
  {journal} {\bibinfo  {journal} {Computational Materials Science}\ }\textbf
  {\bibinfo {volume} {226}},\ \bibinfo {pages} {112230} (\bibinfo {year}
  {2023})}\BibitemShut {NoStop}%
\bibitem [{\citenamefont {Steinhardt}\ \emph {et~al.}(1983)\citenamefont
  {Steinhardt}, \citenamefont {Nelson},\ and\ \citenamefont
  {Ronchetti}}]{SteinhardtPRB1983}%
  \BibitemOpen
  \bibfield  {author} {\bibinfo {author} {\bibfnamefont {P.~J.}\ \bibnamefont
  {Steinhardt}}, \bibinfo {author} {\bibfnamefont {D.~R.}\ \bibnamefont
  {Nelson}},\ and\ \bibinfo {author} {\bibfnamefont {M.}~\bibnamefont
  {Ronchetti}},\ }\bibfield  {title} {\bibinfo {title} {Bond-orientational
  order in liquids and glasses},\ }\href
  {https://doi.org/10.1103/PhysRevB.28.784} {\bibfield  {journal} {\bibinfo
  {journal} {Phys. Rev. B}\ }\textbf {\bibinfo {volume} {28}},\ \bibinfo
  {pages} {784} (\bibinfo {year} {1983})}\BibitemShut {NoStop}%
\bibitem [{\citenamefont {Falk}\ and\ \citenamefont
  {Langer}(1998)}]{FalkPRE1998}%
  \BibitemOpen
  \bibfield  {author} {\bibinfo {author} {\bibfnamefont {M.~L.}\ \bibnamefont
  {Falk}}\ and\ \bibinfo {author} {\bibfnamefont {J.~S.}\ \bibnamefont
  {Langer}},\ }\bibfield  {title} {\bibinfo {title} {Dynamics of viscoplastic
  deformation in amorphous solids},\ }\href
  {https://doi.org/10.1103/PhysRevE.57.7192} {\bibfield  {journal} {\bibinfo
  {journal} {Phys. Rev. E}\ }\textbf {\bibinfo {volume} {57}},\ \bibinfo
  {pages} {7192} (\bibinfo {year} {1998})}\BibitemShut {NoStop}%
\bibitem [{\citenamefont {Kun}\ \emph {et~al.}(2008)\citenamefont {Kun},
  \citenamefont {Carmona}, \citenamefont {Andrade},\ and\ \citenamefont
  {Herrmann}}]{KunPRL2008}%
  \BibitemOpen
  \bibfield  {author} {\bibinfo {author} {\bibfnamefont {F.}~\bibnamefont
  {Kun}}, \bibinfo {author} {\bibfnamefont {H.~A.}\ \bibnamefont {Carmona}},
  \bibinfo {author} {\bibfnamefont {J.~S.}\ \bibnamefont {Andrade}},\ and\
  \bibinfo {author} {\bibfnamefont {H.~J.}\ \bibnamefont {Herrmann}},\
  }\bibfield  {title} {\bibinfo {title} {Universality behind basquin's law of
  fatigue},\ }\href {https://doi.org/10.1103/PhysRevLett.100.094301} {\bibfield
   {journal} {\bibinfo  {journal} {Phys. Rev. Lett.}\ }\textbf {\bibinfo
  {volume} {100}},\ \bibinfo {pages} {094301} (\bibinfo {year}
  {2008})}\BibitemShut {NoStop}%
\bibitem [{\citenamefont {Kurotani}\ and\ \citenamefont
  {Tanaka}(2022)}]{KurotaniCommunMat2022}%
  \BibitemOpen
  \bibfield  {author} {\bibinfo {author} {\bibfnamefont {Y.}~\bibnamefont
  {Kurotani}}\ and\ \bibinfo {author} {\bibfnamefont {H.}~\bibnamefont
  {Tanaka}},\ }\bibfield  {title} {\bibinfo {title} {Fatigue fracture mechanism
  of amorphous materials from a density-based coarse-grained model},\
  }\bibfield  {journal} {\bibinfo  {journal} {Commun Mater}\ }\textbf {\bibinfo
  {volume} {3}},\ \href
  {https://doi.org/https://doi.org/10.1038/s43246-022-00293-9}
  {https://doi.org/10.1038/s43246-022-00293-9} (\bibinfo {year}
  {2022})\BibitemShut {NoStop}%
\bibitem [{\citenamefont {Cabriolu}\ \emph {et~al.}(2019)\citenamefont
  {Cabriolu}, \citenamefont {Horbach}, \citenamefont {Chaudhuri},\ and\
  \citenamefont {Martens}}]{CabrioluSM19}%
  \BibitemOpen
  \bibfield  {author} {\bibinfo {author} {\bibfnamefont {R.}~\bibnamefont
  {Cabriolu}}, \bibinfo {author} {\bibfnamefont {J.}~\bibnamefont {Horbach}},
  \bibinfo {author} {\bibfnamefont {P.}~\bibnamefont {Chaudhuri}},\ and\
  \bibinfo {author} {\bibfnamefont {K.}~\bibnamefont {Martens}},\ }\bibfield
  {title} {\bibinfo {title} {Precursors of fluidisation in the creep response
  of a soft glass},\ }\href {https://doi.org/10.1039/C8SM01432A} {\bibfield
  {journal} {\bibinfo  {journal} {Soft Matter}\ }\textbf {\bibinfo {volume}
  {15}},\ \bibinfo {pages} {415} (\bibinfo {year} {2019})}\BibitemShut
  {NoStop}%
\bibitem [{\citenamefont {Bhaumik}\ \emph
  {et~al.}(2022{\natexlab{a}})\citenamefont {Bhaumik}, \citenamefont {Foffi},\
  and\ \citenamefont {Sastry}}]{BhaumikJCP2022}%
  \BibitemOpen
  \bibfield  {author} {\bibinfo {author} {\bibfnamefont {H.}~\bibnamefont
  {Bhaumik}}, \bibinfo {author} {\bibfnamefont {G.}~\bibnamefont {Foffi}},\
  and\ \bibinfo {author} {\bibfnamefont {S.}~\bibnamefont {Sastry}},\
  }\bibfield  {title} {\bibinfo {title} {{Yielding transition of a two
  dimensional glass former under athermal cyclic shear deformation}},\ }\href
  {https://doi.org/10.1063/5.0085064} {\bibfield  {journal} {\bibinfo
  {journal} {The Journal of Chemical Physics}\ }\textbf {\bibinfo {volume}
  {156}},\ \bibinfo {pages} {064502} (\bibinfo {year}
  {2022}{\natexlab{a}})}\BibitemShut {NoStop}%
\bibitem [{\citenamefont {Sethna}\ \emph {et~al.}(2017)\citenamefont {Sethna},
  \citenamefont {Bierbaum}, \citenamefont {Dahmen}, \citenamefont {Goodrich},
  \citenamefont {Greer}, \citenamefont {Hayden}, \citenamefont {Kent-Dobias},
  \citenamefont {Lee}, \citenamefont {Liarte}, \citenamefont {Ni},
  \citenamefont {Quinn}, \citenamefont {Raju}, \citenamefont {Rocklin},
  \citenamefont {Shekhawat},\ and\ \citenamefont
  {Zapperi}}]{SethnaAnnuRevMaterRes2017}%
  \BibitemOpen
  \bibfield  {author} {\bibinfo {author} {\bibfnamefont {J.~P.}\ \bibnamefont
  {Sethna}}, \bibinfo {author} {\bibfnamefont {M.~K.}\ \bibnamefont
  {Bierbaum}}, \bibinfo {author} {\bibfnamefont {K.~A.}\ \bibnamefont
  {Dahmen}}, \bibinfo {author} {\bibfnamefont {C.~P.}\ \bibnamefont
  {Goodrich}}, \bibinfo {author} {\bibfnamefont {J.~R.}\ \bibnamefont {Greer}},
  \bibinfo {author} {\bibfnamefont {L.~X.}\ \bibnamefont {Hayden}}, \bibinfo
  {author} {\bibfnamefont {J.~P.}\ \bibnamefont {Kent-Dobias}}, \bibinfo
  {author} {\bibfnamefont {E.~D.}\ \bibnamefont {Lee}}, \bibinfo {author}
  {\bibfnamefont {D.~B.}\ \bibnamefont {Liarte}}, \bibinfo {author}
  {\bibfnamefont {X.}~\bibnamefont {Ni}}, \bibinfo {author} {\bibfnamefont
  {K.~N.}\ \bibnamefont {Quinn}}, \bibinfo {author} {\bibfnamefont
  {A.}~\bibnamefont {Raju}}, \bibinfo {author} {\bibfnamefont {D.~Z.}\
  \bibnamefont {Rocklin}}, \bibinfo {author} {\bibfnamefont {A.}~\bibnamefont
  {Shekhawat}},\ and\ \bibinfo {author} {\bibfnamefont {S.}~\bibnamefont
  {Zapperi}},\ }\bibfield  {title} {\bibinfo {title} {Deformation of crystals:
  Connections with statistical physics},\ }\href
  {https://doi.org/10.1146/annurev-matsci-070115-032036} {\bibfield  {journal}
  {\bibinfo  {journal} {Annual Review of Materials Research}\ }\textbf
  {\bibinfo {volume} {47}},\ \bibinfo {pages} {217} (\bibinfo {year}
  {2017})}\BibitemShut {NoStop}%
\bibitem [{\citenamefont {Bhaumik}\ \emph
  {et~al.}(2022{\natexlab{b}})\citenamefont {Bhaumik}, \citenamefont {Foffi},\
  and\ \citenamefont {Sastry}}]{BhaumikPRL2022}%
  \BibitemOpen
  \bibfield  {author} {\bibinfo {author} {\bibfnamefont {H.}~\bibnamefont
  {Bhaumik}}, \bibinfo {author} {\bibfnamefont {G.}~\bibnamefont {Foffi}},\
  and\ \bibinfo {author} {\bibfnamefont {S.}~\bibnamefont {Sastry}},\
  }\bibfield  {title} {\bibinfo {title} {Avalanches, clusters, and structural
  change in cyclically sheared silica glass},\ }\href
  {https://doi.org/10.1103/PhysRevLett.128.098001} {\bibfield  {journal}
  {\bibinfo  {journal} {Phys. Rev. Lett.}\ }\textbf {\bibinfo {volume} {128}},\
  \bibinfo {pages} {098001} (\bibinfo {year} {2022}{\natexlab{b}})}\BibitemShut
  {NoStop}%
\bibitem [{\citenamefont {Richard}\ \emph {et~al.}(2020)\citenamefont
  {Richard}, \citenamefont {Ozawa}, \citenamefont {Patinet}, \citenamefont
  {Stanifer}, \citenamefont {Shang}, \citenamefont {Ridout}, \citenamefont
  {Xu}, \citenamefont {Zhang}, \citenamefont {Morse}, \citenamefont {Barrat},
  \citenamefont {Berthier}, \citenamefont {Falk}, \citenamefont {Guan},
  \citenamefont {Liu}, \citenamefont {Martens}, \citenamefont {Sastry},
  \citenamefont {Vandembroucq}, \citenamefont {Lerner},\ and\ \citenamefont
  {Manning}}]{Richard_PhysRevMaterials.4.113609}%
  \BibitemOpen
  \bibfield  {author} {\bibinfo {author} {\bibfnamefont {D.}~\bibnamefont
  {Richard}}, \bibinfo {author} {\bibfnamefont {M.}~\bibnamefont {Ozawa}},
  \bibinfo {author} {\bibfnamefont {S.}~\bibnamefont {Patinet}}, \bibinfo
  {author} {\bibfnamefont {E.}~\bibnamefont {Stanifer}}, \bibinfo {author}
  {\bibfnamefont {B.}~\bibnamefont {Shang}}, \bibinfo {author} {\bibfnamefont
  {S.~A.}\ \bibnamefont {Ridout}}, \bibinfo {author} {\bibfnamefont
  {B.}~\bibnamefont {Xu}}, \bibinfo {author} {\bibfnamefont {G.}~\bibnamefont
  {Zhang}}, \bibinfo {author} {\bibfnamefont {P.~K.}\ \bibnamefont {Morse}},
  \bibinfo {author} {\bibfnamefont {J.-L.}\ \bibnamefont {Barrat}}, \bibinfo
  {author} {\bibfnamefont {L.}~\bibnamefont {Berthier}}, \bibinfo {author}
  {\bibfnamefont {M.~L.}\ \bibnamefont {Falk}}, \bibinfo {author}
  {\bibfnamefont {P.}~\bibnamefont {Guan}}, \bibinfo {author} {\bibfnamefont
  {A.~J.}\ \bibnamefont {Liu}}, \bibinfo {author} {\bibfnamefont
  {K.}~\bibnamefont {Martens}}, \bibinfo {author} {\bibfnamefont
  {S.}~\bibnamefont {Sastry}}, \bibinfo {author} {\bibfnamefont
  {D.}~\bibnamefont {Vandembroucq}}, \bibinfo {author} {\bibfnamefont
  {E.}~\bibnamefont {Lerner}},\ and\ \bibinfo {author} {\bibfnamefont {M.~L.}\
  \bibnamefont {Manning}},\ }\bibfield  {title} {\bibinfo {title} {Predicting
  plasticity in disordered solids from structural indicators},\ }\href
  {https://doi.org/10.1103/PhysRevMaterials.4.113609} {\bibfield  {journal}
  {\bibinfo  {journal} {Phys. Rev. Mater.}\ }\textbf {\bibinfo {volume} {4}},\
  \bibinfo {pages} {113609} (\bibinfo {year} {2020})}\BibitemShut {NoStop}%
\bibitem [{\citenamefont {Lerner}\ and\ \citenamefont
  {Bouchbinder}(2021)}]{Lerner2021}%
  \BibitemOpen
  \bibfield  {author} {\bibinfo {author} {\bibfnamefont {E.}~\bibnamefont
  {Lerner}}\ and\ \bibinfo {author} {\bibfnamefont {E.}~\bibnamefont
  {Bouchbinder}},\ }\bibfield  {title} {\bibinfo {title} {{Low-energy
  quasilocalized excitations in structural glasses}},\ }\href
  {https://doi.org/10.1063/5.0069477} {\bibfield  {journal} {\bibinfo
  {journal} {The Journal of Chemical Physics}\ }\textbf {\bibinfo {volume}
  {155}},\ \bibinfo {pages} {200901} (\bibinfo {year} {2021})}\BibitemShut
  {NoStop}%
\bibitem [{\citenamefont {Wu}\ \emph {et~al.}(2023)\citenamefont {Wu},
  \citenamefont {Chen}, \citenamefont {Wang}, \citenamefont {Kob},\ and\
  \citenamefont {Xu}}]{WuNatCom2023}%
  \BibitemOpen
  \bibfield  {author} {\bibinfo {author} {\bibfnamefont {Z.~W.}\ \bibnamefont
  {Wu}}, \bibinfo {author} {\bibfnamefont {Y.}~\bibnamefont {Chen}}, \bibinfo
  {author} {\bibfnamefont {W.-H.}\ \bibnamefont {Wang}}, \bibinfo {author}
  {\bibfnamefont {W.}~\bibnamefont {Kob}},\ and\ \bibinfo {author}
  {\bibfnamefont {L.}~\bibnamefont {Xu}},\ }\bibfield  {title} {\bibinfo
  {title} {Topology of vibrational modes predicts plastic events in glasses},\
  }\href {https://doi.org/10.1038/s41467-023-38547-w} {\bibfield  {journal}
  {\bibinfo  {journal} {Nature Communications}\ }\textbf {\bibinfo {volume}
  {14}},\ \bibinfo {pages} {2955} (\bibinfo {year} {2023})}\BibitemShut
  {NoStop}%
\bibitem [{\citenamefont {Desmarchelier}\ \emph {et~al.}(2024)\citenamefont
  {Desmarchelier}, \citenamefont {Fajardo},\ and\ \citenamefont
  {Falk}}]{Falk_2024_PhysRevE.109.L053002}%
  \BibitemOpen
  \bibfield  {author} {\bibinfo {author} {\bibfnamefont {P.}~\bibnamefont
  {Desmarchelier}}, \bibinfo {author} {\bibfnamefont {S.}~\bibnamefont
  {Fajardo}},\ and\ \bibinfo {author} {\bibfnamefont {M.~L.}\ \bibnamefont
  {Falk}},\ }\bibfield  {title} {\bibinfo {title} {Topological characterization
  of rearrangements in amorphous solids},\ }\href
  {https://doi.org/10.1103/PhysRevE.109.L053002} {\bibfield  {journal}
  {\bibinfo  {journal} {Phys. Rev. E}\ }\textbf {\bibinfo {volume} {109}},\
  \bibinfo {pages} {L053002} (\bibinfo {year} {2024})}\BibitemShut {NoStop}%
\bibitem [{\citenamefont {Bera}\ \emph {et~al.}(2024)\citenamefont {Bera},
  \citenamefont {Baggioli}, \citenamefont {Petersen}, \citenamefont {Sirk},
  \citenamefont {Liu},\ and\ \citenamefont {Zaccone}}]{Zaccone_PNASnexus24}%
  \BibitemOpen
  \bibfield  {author} {\bibinfo {author} {\bibfnamefont {A.}~\bibnamefont
  {Bera}}, \bibinfo {author} {\bibfnamefont {M.}~\bibnamefont {Baggioli}},
  \bibinfo {author} {\bibfnamefont {T.~C.}\ \bibnamefont {Petersen}}, \bibinfo
  {author} {\bibfnamefont {T.~W.}\ \bibnamefont {Sirk}}, \bibinfo {author}
  {\bibfnamefont {A.~C.~Y.}\ \bibnamefont {Liu}},\ and\ \bibinfo {author}
  {\bibfnamefont {A.}~\bibnamefont {Zaccone}},\ }\bibfield  {title} {\bibinfo
  {title} {{Clustering of negative topological charges precedes plastic failure
  in 3D glasses}},\ }\href {https://doi.org/10.1093/pnasnexus/pgae315}
  {\bibfield  {journal} {\bibinfo  {journal} {PNAS Nexus}\ }\textbf {\bibinfo
  {volume} {3}},\ \bibinfo {pages} {pgae315} (\bibinfo {year}
  {2024})}\BibitemShut {NoStop}%
\bibitem [{\citenamefont {Chikkadi}\ \emph {et~al.}(2015)\citenamefont
  {Chikkadi}, \citenamefont {Gendelman}, \citenamefont {Ilyin}, \citenamefont
  {Ashwin}, \citenamefont {Procaccia},\ and\ \citenamefont
  {Shor}}]{Chikkadi_2015}%
  \BibitemOpen
  \bibfield  {author} {\bibinfo {author} {\bibfnamefont {V.}~\bibnamefont
  {Chikkadi}}, \bibinfo {author} {\bibfnamefont {O.}~\bibnamefont {Gendelman}},
  \bibinfo {author} {\bibfnamefont {V.}~\bibnamefont {Ilyin}}, \bibinfo
  {author} {\bibfnamefont {J.}~\bibnamefont {Ashwin}}, \bibinfo {author}
  {\bibfnamefont {I.}~\bibnamefont {Procaccia}},\ and\ \bibinfo {author}
  {\bibfnamefont {C.~A. B.~Z.}\ \bibnamefont {Shor}},\ }\bibfield  {title}
  {\bibinfo {title} {Spreading plastic failure as a mechanism for the shear
  modulus reduction in amorphous solids},\ }\href
  {https://doi.org/10.1209/0295-5075/110/48001} {\bibfield  {journal} {\bibinfo
   {journal} {Europhysics Letters}\ }\textbf {\bibinfo {volume} {110}},\
  \bibinfo {pages} {48001} (\bibinfo {year} {2015})}\BibitemShut {NoStop}%
\bibitem [{\citenamefont {Shrivastav}\ \emph {et~al.}(2016)\citenamefont
  {Shrivastav}, \citenamefont {Chaudhuri},\ and\ \citenamefont
  {Horbach}}]{Shrivastav2016}%
  \BibitemOpen
  \bibfield  {author} {\bibinfo {author} {\bibfnamefont {G.~P.}\ \bibnamefont
  {Shrivastav}}, \bibinfo {author} {\bibfnamefont {P.}~\bibnamefont
  {Chaudhuri}},\ and\ \bibinfo {author} {\bibfnamefont {J.}~\bibnamefont
  {Horbach}},\ }\bibfield  {title} {\bibinfo {title} {Yielding of glass under
  shear: A directed percolation transition precedes shear-band formation},\
  }\href {https://doi.org/10.1103/PhysRevE.94.042605} {\bibfield  {journal}
  {\bibinfo  {journal} {Phys. Rev. E}\ }\textbf {\bibinfo {volume} {94}},\
  \bibinfo {pages} {042605} (\bibinfo {year} {2016})}\BibitemShut {NoStop}%
\bibitem [{\citenamefont {Ghosh}\ \emph {et~al.}(2017)\citenamefont {Ghosh},
  \citenamefont {Budrikis}, \citenamefont {Chikkadi}, \citenamefont {Sellerio},
  \citenamefont {Zapperi},\ and\ \citenamefont {Schall}}]{GhoshPRL2017}%
  \BibitemOpen
  \bibfield  {author} {\bibinfo {author} {\bibfnamefont {A.}~\bibnamefont
  {Ghosh}}, \bibinfo {author} {\bibfnamefont {Z.}~\bibnamefont {Budrikis}},
  \bibinfo {author} {\bibfnamefont {V.}~\bibnamefont {Chikkadi}}, \bibinfo
  {author} {\bibfnamefont {A.~L.}\ \bibnamefont {Sellerio}}, \bibinfo {author}
  {\bibfnamefont {S.}~\bibnamefont {Zapperi}},\ and\ \bibinfo {author}
  {\bibfnamefont {P.}~\bibnamefont {Schall}},\ }\bibfield  {title} {\bibinfo
  {title} {Direct observation of percolation in the yielding transition of
  colloidal glasses},\ }\href {https://doi.org/10.1103/PhysRevLett.118.148001}
  {\bibfield  {journal} {\bibinfo  {journal} {Phys. Rev. Lett.}\ }\textbf
  {\bibinfo {volume} {118}},\ \bibinfo {pages} {148001} (\bibinfo {year}
  {2017})}\BibitemShut {NoStop}%
\bibitem [{\citenamefont {\ifmmode~\mbox{\c{S}}\else \c{S}\fi{}opu}\ \emph
  {et~al.}(2017)\citenamefont {\ifmmode~\mbox{\c{S}}\else \c{S}\fi{}opu},
  \citenamefont {Stukowski}, \citenamefont {Stoica},\ and\ \citenamefont
  {Scudino}}]{SopuPRL2017}%
  \BibitemOpen
  \bibfield  {author} {\bibinfo {author} {\bibfnamefont {D.}~\bibnamefont
  {\ifmmode~\mbox{\c{S}}\else \c{S}\fi{}opu}}, \bibinfo {author} {\bibfnamefont
  {A.}~\bibnamefont {Stukowski}}, \bibinfo {author} {\bibfnamefont
  {M.}~\bibnamefont {Stoica}},\ and\ \bibinfo {author} {\bibfnamefont
  {S.}~\bibnamefont {Scudino}},\ }\bibfield  {title} {\bibinfo {title}
  {Atomic-level processes of shear band nucleation in metallic glasses},\
  }\href {https://doi.org/10.1103/PhysRevLett.119.195503} {\bibfield  {journal}
  {\bibinfo  {journal} {Phys. Rev. Lett.}\ }\textbf {\bibinfo {volume} {119}},\
  \bibinfo {pages} {195503} (\bibinfo {year} {2017})}\BibitemShut {NoStop}%
\bibitem [{\citenamefont {Das}\ \emph {et~al.}(2022)\citenamefont {Das},
  \citenamefont {Parmar},\ and\ \citenamefont {Sastry}}]{PallabiJCP2022}%
  \BibitemOpen
  \bibfield  {author} {\bibinfo {author} {\bibfnamefont {P.}~\bibnamefont
  {Das}}, \bibinfo {author} {\bibfnamefont {A.~D.~S.}\ \bibnamefont {Parmar}},\
  and\ \bibinfo {author} {\bibfnamefont {S.}~\bibnamefont {Sastry}},\
  }\bibfield  {title} {\bibinfo {title} {{Annealing glasses by cyclic shear
  deformation}},\ }\href {https://doi.org/10.1063/5.0100523} {\bibfield
  {journal} {\bibinfo  {journal} {The Journal of Chemical Physics}\ }\textbf
  {\bibinfo {volume} {157}},\ \bibinfo {pages} {044501} (\bibinfo {year}
  {2022})}\BibitemShut {NoStop}%
\bibitem [{\citenamefont {Evans}\ and\ \citenamefont
  {Morriss}(1984)}]{EvansPRA1984}%
  \BibitemOpen
  \bibfield  {author} {\bibinfo {author} {\bibfnamefont {D.~J.}\ \bibnamefont
  {Evans}}\ and\ \bibinfo {author} {\bibfnamefont {G.~P.}\ \bibnamefont
  {Morriss}},\ }\bibfield  {title} {\bibinfo {title} {Nonlinear-response theory
  for steady planar couette flow},\ }\href
  {https://doi.org/10.1103/PhysRevA.30.1528} {\bibfield  {journal} {\bibinfo
  {journal} {Phys. Rev. A}\ }\textbf {\bibinfo {volume} {30}},\ \bibinfo
  {pages} {1528} (\bibinfo {year} {1984})}\BibitemShut {NoStop}%
\bibitem [{\citenamefont {Thompson}\ \emph {et~al.}(2022)\citenamefont
  {Thompson}, \citenamefont {Aktulga}, \citenamefont {Berger}, \citenamefont
  {Bolintineanu}, \citenamefont {Brown}, \citenamefont {Crozier}, \citenamefont
  {in~'t Veld}, \citenamefont {Kohlmeyer}, \citenamefont {Moore}, \citenamefont
  {Nguyen}, \citenamefont {Shan}, \citenamefont {Stevens}, \citenamefont
  {Tranchida}, \citenamefont {Trott},\ and\ \citenamefont {Plimpton}}]{LAMMPS}%
  \BibitemOpen
  \bibfield  {author} {\bibinfo {author} {\bibfnamefont {A.~P.}\ \bibnamefont
  {Thompson}}, \bibinfo {author} {\bibfnamefont {H.~M.}\ \bibnamefont
  {Aktulga}}, \bibinfo {author} {\bibfnamefont {R.}~\bibnamefont {Berger}},
  \bibinfo {author} {\bibfnamefont {D.~S.}\ \bibnamefont {Bolintineanu}},
  \bibinfo {author} {\bibfnamefont {W.~M.}\ \bibnamefont {Brown}}, \bibinfo
  {author} {\bibfnamefont {P.~S.}\ \bibnamefont {Crozier}}, \bibinfo {author}
  {\bibfnamefont {P.~J.}\ \bibnamefont {in~'t Veld}}, \bibinfo {author}
  {\bibfnamefont {A.}~\bibnamefont {Kohlmeyer}}, \bibinfo {author}
  {\bibfnamefont {S.~G.}\ \bibnamefont {Moore}}, \bibinfo {author}
  {\bibfnamefont {T.~D.}\ \bibnamefont {Nguyen}}, \bibinfo {author}
  {\bibfnamefont {R.}~\bibnamefont {Shan}}, \bibinfo {author} {\bibfnamefont
  {M.~J.}\ \bibnamefont {Stevens}}, \bibinfo {author} {\bibfnamefont
  {J.}~\bibnamefont {Tranchida}}, \bibinfo {author} {\bibfnamefont
  {C.}~\bibnamefont {Trott}},\ and\ \bibinfo {author} {\bibfnamefont {S.~J.}\
  \bibnamefont {Plimpton}},\ }\bibfield  {title} {\bibinfo {title} {{LAMMPS} -
  a flexible simulation tool for particle-based materials modeling at the
  atomic, meso, and continuum scales},\ }\href
  {https://doi.org/10.1016/j.cpc.2021.108171} {\bibfield  {journal} {\bibinfo
  {journal} {Comp. Phys. Comm.}\ }\textbf {\bibinfo {volume} {271}},\ \bibinfo
  {pages} {108171} (\bibinfo {year} {2022})}\BibitemShut {NoStop}%
\end{thebibliography}%

\clearpage

\noindent{\bf{\large Methods:}}

\noindent{\bf {\em Model and Simulation details:}}
We simulate the 80:20 Kob-Anderson binary mixture of Lennard-Jones particles interacting via the potential \cite{LeishanthemNatCom2017}
\begin{equation}
\begin{aligned}
    U_{\alpha \beta}(r) =& 4\epsilon_{\alpha\beta}\left[ \left(\frac{\sigma_{\alpha\beta}}{r}\right)^{12} - \left(\frac{\sigma_{\alpha\beta}}{r}\right)^{6} \right]\\
    &+ 4\epsilon_{\alpha\beta}\left[ c_{0\alpha\beta} + c_{2\alpha\beta}\left(\frac{r}{\sigma_{\alpha\beta}}\right)^2 \right]\text{, for } r<r_{c\alpha\beta}\\
    =&0 \text{, for } r>r_{c\alpha\beta},
\end{aligned}
\end{equation}
where $\alpha,\beta \in A,B$ are particle types with the parameters $\epsilon_{AB}/\epsilon_{AA} = 1.5$, $\epsilon_{BB}/\epsilon_{AA} = 0.5$, $\sigma_{AB}/\sigma_{AA} = 0.8$, $\sigma_{BB}/\sigma_{AA} = 0.88$, $r_{c\alpha\beta} = 2.5\sigma_{\alpha\beta}$. $c_{0\alpha\beta}$, $c_{2\alpha\beta}$ are chosen such that both the potential and the force go to zero at $r_{c\alpha\beta}$: $c_{0\alpha\beta} = -7\left(\frac{\sigma_{\alpha\beta}}{r_{c\alpha\beta}}\right)^{12} + 4\left(\frac{\sigma_{\alpha\beta}}{r_{c\alpha\beta}}\right)^{6}$ and $c_{2\alpha\beta} = 6\left(\frac{\sigma_{\alpha\beta}}{r_{c\alpha\beta}}\right)^{14} - 3\left(\frac{\sigma_{\alpha\beta}}{r_{c\alpha\beta}}\right)^{8}$. Units of length, energy, and time are chosen to be $\sigma_{AA}, \epsilon_{AA}$ and $\sqrt{\frac{\sigma_{AA}^2}{\epsilon_{AA}}}$ respectively. We use an integration time step of $0.005$ equilibration. 


\noindent{\bf {\em Initial glass preparation:}} We performed constant
volume, temperature (NVT) molecular dynamics simulations using the Nos{\'e}-Hoover thermostat for a wide range of parent $T_p\in [0.435-1]$ and reduced density $1.2$. Configurations sampled from equilibrated trajectories are subjected to local energy minimization to obtain the so-called inherent structures, which we employ as the glass configurations at zero temperature. Lower parent temperatures correspond to better annealed glasses. Glasses obtain in this manner have per particle energies $e_{IS} >-7.00$. We also obtain lower energy inherent structures (those have energy $e_{IS}<-7.00$) with the method reported in Ref.\cite{PallabiJCP2022}. Most of the results here are reported for  $e_{IS}=-6.96$, and $-7.00$, which we call as poorly annealed and well-annealed glasses respectively \cite{BhaumikPNAS2021}.

\noindent{\bf {\em Cyclic shear simulations:}}
We perform non-equilibrium molecular dynamics simulations to implement the strain-controlled cyclic shear at finite temperatures and strain rates, by solving the Nos\'e-Hoover thermostatted SLLOD\cite{EvansPRA1984} equation motion\:
\begin{align}
    \dot{\mathbf{r}}_i&=\frac{\mathbf{p}_i}{m} + \dot{\gamma}_{xy}\mathbf{r}_{yi}\hat{i}\nonumber\\
    \dot{\mathbf{p}}_i&=\frac{\mathbf{F}_i}{m} - \dot{\gamma}_{xy}\mathbf{p}_{yi}\hat{i}-\alpha(t)\mathbf{p}_i
\end{align}
The strain $\gamma_{xy}$ of the system is varied as $\gamma_{xy}(t) = \gamma_{max} \sin(\omega t)$, where $\gamma_{max}$ is the maximum strain amplitude of the cycles. The shear rate at the start of the cycles is taken to be $\Dot{\gamma}_{xy}(t=0)=\gamma_{max}\omega = 0.001$. We choose the thermostat temperature to be $T=0.001$. For cyclic shear, we choose integration time step 0.01. The damping coefficient $\alpha(t)$ for the Nos\'e-Hoover thermostat is $\alpha(t)=p_\eta(t)/Q$, where $p_\eta$ is momentum for thermostat coordinate $\eta$ which evolves as:
\begin{equation}
    \dot{p}_\eta=\sum_i^N\mathbf{p}_i^2/m-N_f k_B T
\end{equation}
where $N_f$ is the number of degree of freedoms. $Q$ is the mass of the thermostat which is associated with thermostat relaxation time $\tau$ as: $Q=N_f k_B T \tau^2$. We have chosen to be $\tau=1.0$. For the small systems of size $N=4000$, at least $60$ independent samples are studied for each $\gamma_{max}$ ranging from $0.05-0.14$ for both degrees of annealing. For larger system sizes of $N=64000$ and $N=128000$, $32$ samples are studied for each strain amplitude and each degree of annealing mentioned in the text.
All the numerical simulations including molecular dynamics, cyclic shear, and energy minimization via conjugate-gradient algorithm, are performed using LAMMPS\cite{LAMMPS}.


\noindent{\bf {\em Definition of metrics to identify failure}}
To estimate the failure time, we consider various structural and mechanical properties of the system {\em e.g} per particle potential energy($U/N$), bond orientational order parameter $q_6$, and non-affine displacement $D^2_{min}$. All these metrics are per particle quantities and we report averages over all the particles. In the following, we provide a detailed definition of them.

{{\em  Bond orientational order parameter:}}
The bond orientational order parameter $q_l$ is calculated following \cite{SteinhardtPRB1983}:
\begin{align}
    q_{l}(i)=\sqrt{\frac{4\pi}{2l+1}\sum_{m=-l}^{l}\left|q_{lm}(i)\right|^2},\\
    q_{lm}(i)=\frac{1}{N_b(i)}\sum_{j=1}^{N_b(i)}Y_{lm}(\mathbf{r}_{ij}),
\end{align}
where $Y_{lm}$ are spherical harmonics and $N_b(i)$ is the number of neighbours of reference particle $i$. We consider $l=6$. We compute the quantity for  $A$ type of particles, where the neighbor distance is taken to be $1.42$ and $1.26$ for $AA$ and $AB$-type of interactions, respectively. These cutoffs correspond to the 1st minima in the pair correlation function.

{\em Non-affine displacement:} The non-affine displacement $D^2_{min}$ of a particle $i$ from time $t-\Delta t$ to $t$ is calculated as \cite{FalkPRE1998}:
\begin{eqnarray}
D^2_{min}(i)(t,\Delta t)&\!=\!&\sum_j\sum_\alpha  \biggl ( r^\alpha_j(t)-r_i^\alpha(t)-\sum_\beta(\delta_{\alpha\beta}+\epsilon_{\alpha\beta}) \nonumber\\
&&  \times\left[r_j^\beta(t-\Delta t)-r_i^\beta(t-\Delta t) \right] \biggr) ^2 
\end{eqnarray}
where the indices $\alpha$ and $\beta$ denote spatial coordinates and the index $j$ runs over the \textit{neighbour} atoms of the reference atom $i$. Here, the \textit{neighbor} of an atom is defined as a particle that are within a particular distance from the reference particle. For the studied system this distance is taken as $2.5$, $2.0$, and $2.2$ for $AA$, $AB$, and $BB$ pairs respectively. The local strain $\epsilon_{\alpha\beta}$ which minimizes $D^2_{min}$, is calculated as 
\begin{align}
   \epsilon_{\alpha\beta}&=\sum_\gamma X_{\alpha\gamma} Y_{\gamma\beta}^{-1} - \delta_{\alpha\beta},
\end{align}
where
    \begin{align}
    X_{\alpha\beta}=&\sum_j\left[r^\alpha_j(t)-r^\alpha_i(t)\right]\times\left[r^\beta_j(t-\Delta t)-r^\beta_i(t-\Delta t)\right],\nonumber\\ 
    Y_{\alpha\beta}=&\sum_j\left[r^\alpha_j(t-\Delta t)-r^\alpha_i(t-\Delta t)\right]\times\nonumber\\
    & \left[r^\beta_j(t-\Delta t)-r^\beta_i(t-\Delta t)\right].\nonumber
\end{align}

\noindent{\bf {\em Identification of mobile particles:}} To identify the plastically rearranged mobile particle, we compute the cycle to cycle nonaffine displacement $D^{2,cyc}_{min}$ of each particle between two stroboscopic configurations that are $10$ strain cycle apart. The distribution of $D^{2,cyc}_{min}$ exhibits power-law with an exponential tail. We set a cutoff at $D^{2,cyc}_{min} = 1.2$, where the exponential tail begins, and consider particles with values beyond this threshold as mobile particles. This is not yet a very robust procedure, as discussed in the main text, and there is a need for improved criteria.


\noindent{\bf {\em Percolation analysis:}} We perform percolation analysis of accumulated mobile particles\cite{LeishanthemNatCom2017} up to a certain number of strain cycles. In this analysis, we consider two mobile particles to belong to the same cluster if they are within the distance of $1.4$, the first coordination shell of the particles. We extract the largest cluster $S_{max}$ and check further if it is a system-spanning one. To identify that we replicate the simulation box three times in all spatial directions so that the replicated system will have $27$ copies of the original simulation box. We perform cluster computing on this new replicated system. The largest cluster in the replicated system will be three, nine, or $27$ times of $S_{max}$ of the original box if $S_{max}$ in the original box percolates in one, two, or three spatial directions, respectively.



 \noindent{\bf{\large Acknowledgements}}
We thank P. Sollich, I. Procaccia, E. Lerner, A. Zaconne, J. Horbach, M Adhikari, Pushkar Khandare, and Debargha Sarkar for useful discussions and comments on the manuscript. We acknowledge the  National Supercomputing Mission facility (Param Yukti) at the Jawaharlal Nehru Center for Advanced Scientific Research for computational resources. H.B. acknowledges EPSRC for support through grants EP/T031247/1 during a part of the period when this work was performed.  S. S. acknowledges SERB (India) for support through the JC Bose Fellowship (JBR/2020/000015) SERB, DST (India) and SUPRA project number SPR/2021/000382. 









\clearpage
\newpage
\onecolumngrid

\section*{Supplementary Information}{\label{sec:si}}
\renewcommand{\theequation}{S\arabic{equation}}
\renewcommand{\thefigure}{S\arabic{figure}}
\renewcommand{\thesection}{S-\arabic{section}} 
\renewcommand{\thesubsection}{\Alph}
\renewcommand{\thesubsection}{S-\Alph{subsection}}
\renewcommand{\thepage}{S\arabic{page}} 
\setcounter{section}{0} 
\setcounter{subsection}{0} 
\setcounter{figure}{0} 
\setcounter{equation}{0} 
\setcounter{page}{1}

\section*{}
In the supplementary information, we present additional analysis to support our findings. This includes details of the procedure for estimating the yield strain amplitude. We also present examples of fitting through energy, structure, and plasticity data with a sigmoid function for different system sizes, degrees of annealing, and strain amplitudes. We provide details of estimating the failure times from sample-to-sample fluctuations. The dependence of energy on time scaled to the failure time is shown. We provide results on the duration of the failure process, or the transformation width. Finally, we provide data regarding the correlation between plasticity and failure for a poorly annealed glass.

\section{Estimation of the yield strain amplitude} \label{SI_gammaY}

\begin{figure}[!ht]
          \centerline{
            \includegraphics[width=.66\textwidth]{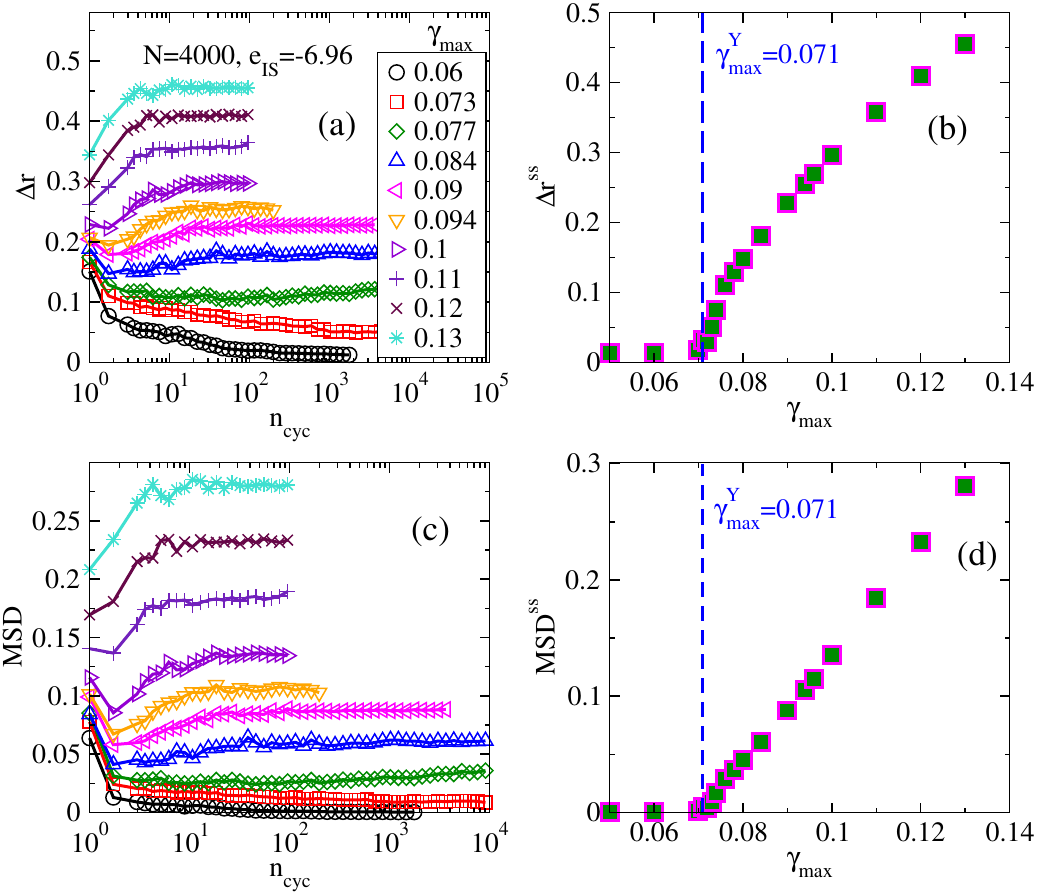}
        }
    \caption{\textbf{Calculation of yield strain amplitude.}  Cycle to cycle (a) Mean displacement $\Delta r$ and (c) Mean squared displacement MSD as a function of strain cycle for $N=4000$ for $e_{IS}=-6.96$. The corresponding steady-state values against strain amplitude $\gamma_{max}$ in (b) and (d), respectively. The yield strain amplitude $\gamma_{max}^Y$, is found to be $0.071\pm 0.001$, as indicated by the vertical dashed lines.}
    \label{fig:yield_strain_calculation_N4K_dr_msd}
\end{figure}

To observe the system's dynamical evolution, we consider per particle cycle-to-cycle displacement $\Delta r$ and cycle-to-cycle mean square displacement $MSD$ as a function of the strain cycle. These quantities are defined as 
\begin{align}
        \Delta r (t)=\frac{1}{N}\sum_{i=1}^{N}\sqrt{\left[\mathbf{r}_i(t)-\mathbf{r}_i(t-1)\right]^2},\\
        MSD(t) =\frac{1}{N}\sum_{i=1}^{N}\left(\mathbf{r}_i(t)-\mathbf{r}_i(t-1)\right)^2,
\end{align}
where $\mathbf{r}_i(t)$ is the position vector of  $i^{th}$ particle after at $t^{th}$ strain cycle.

In Figs. \ref{fig:yield_strain_calculation_N4K_dr_msd} (a) and (c), we show the variation of $\Delta r$ and $MSD$ as a function of strain cycle $n_{cyc}$ for different strain amplitudes $\gamma_{max}$. For small $\gamma_{max}$, $\Delta r$ and $MSD$ decrease with the number of cycles, as the system approaches a limit cycle. For large $\gamma_{max}$,  the evolution is non-monotonic, and both quantities reach a finite steady-state value. This non-monotonic behavior is consistent with observations in previous literature \cite{ParleyPRL2022,ParmarPRX2019}. 

The corresponding steady-state values of $\Delta r^{ss}$ and $MSD^{ss}$ are shown against $\gamma_{max}$ in Fig. \ref{fig:yield_strain_calculation_N4K_dr_msd} (b) and (d) respectively.
Yield strain amplitude $\gamma_{max}^Y$ is identified at the $\gamma_{max}$ where there is a sharp increase in $\Delta r^{ss}$ and $MSD^{ss}$, which is found to be $\gamma_{max}^Y=0.071\pm0.001$, for the system size of $N=4000$. 

It should be noted that the value of $\gamma_{max}^Y$ is sensitive to the system size and degree of annealing. We use the above method to extract $\gamma_{max}^Y$ for different degrees of annealing and system sizes we consider in this study.

\section{Calculation of Failure times}  \label{SI_tf}

\begin{figure}[ht!]
        \centering
        {\includegraphics[width=.22\textwidth]{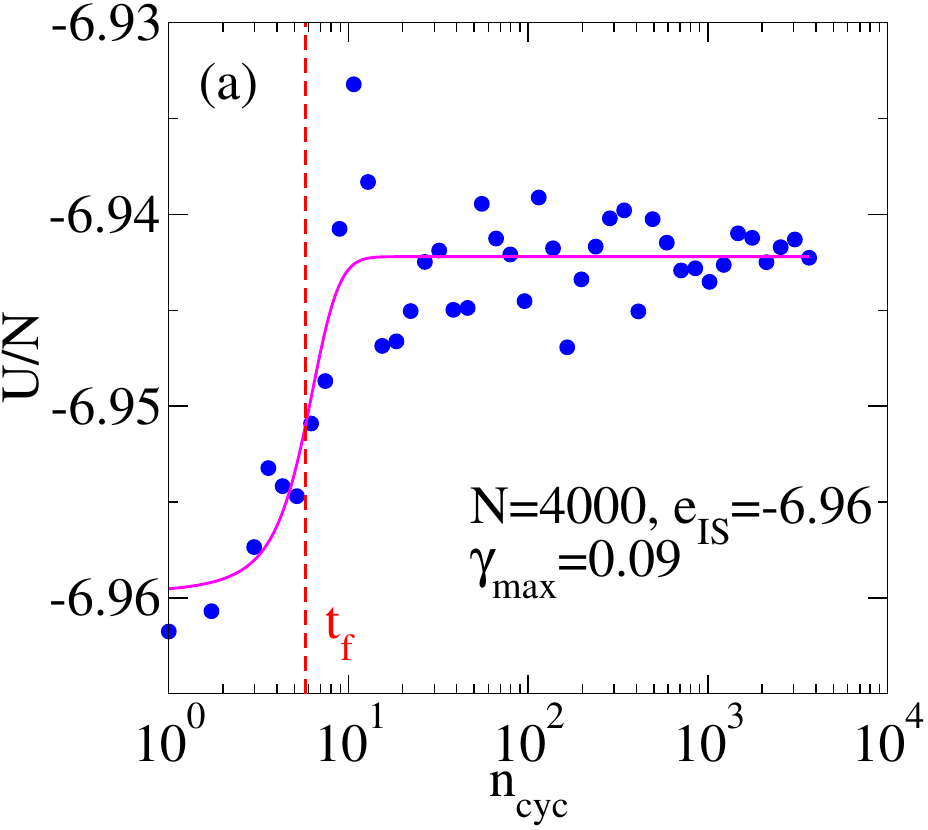}}
        {\includegraphics[width=.22\textwidth]{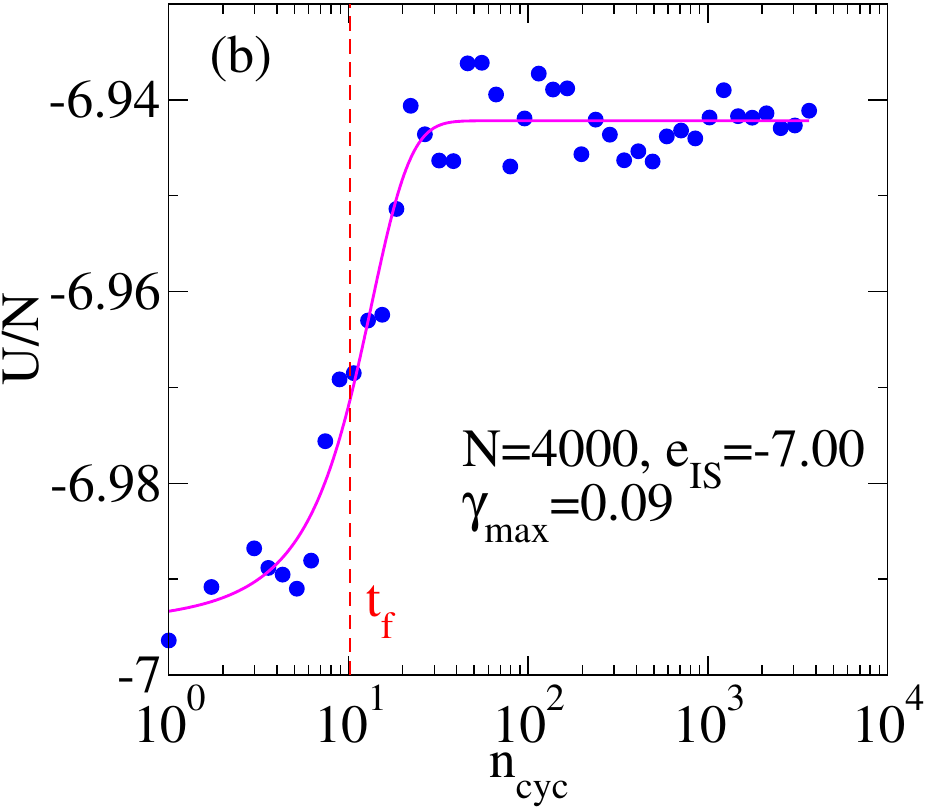}}
        {\includegraphics[width=.23\textwidth]{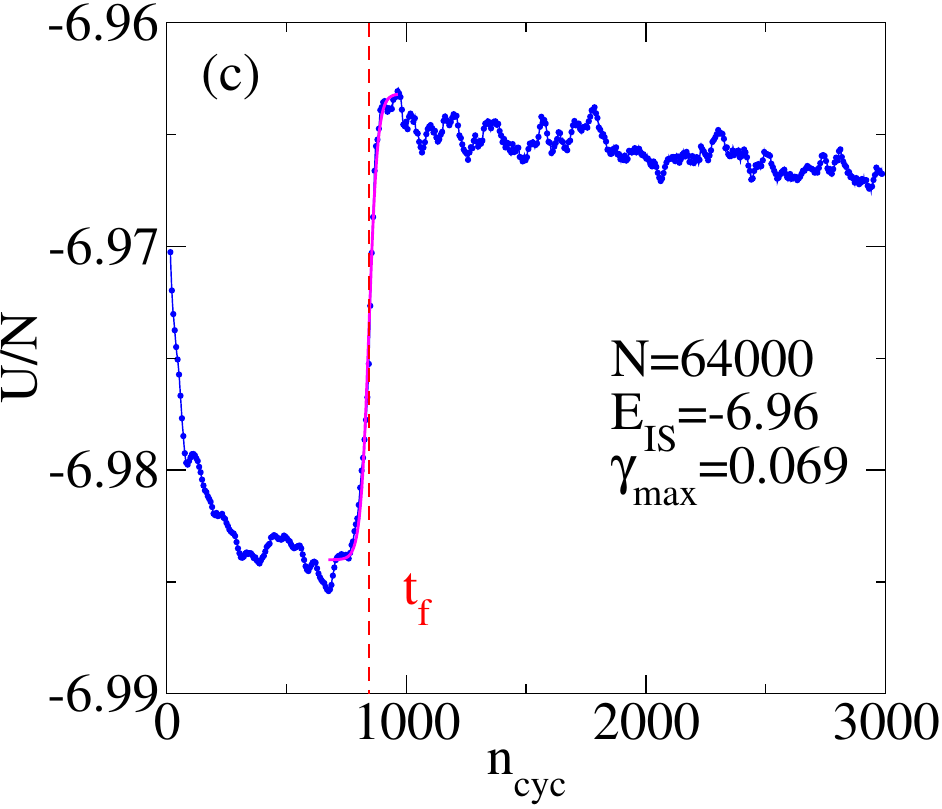}}
        {\includegraphics[width=.23\textwidth]{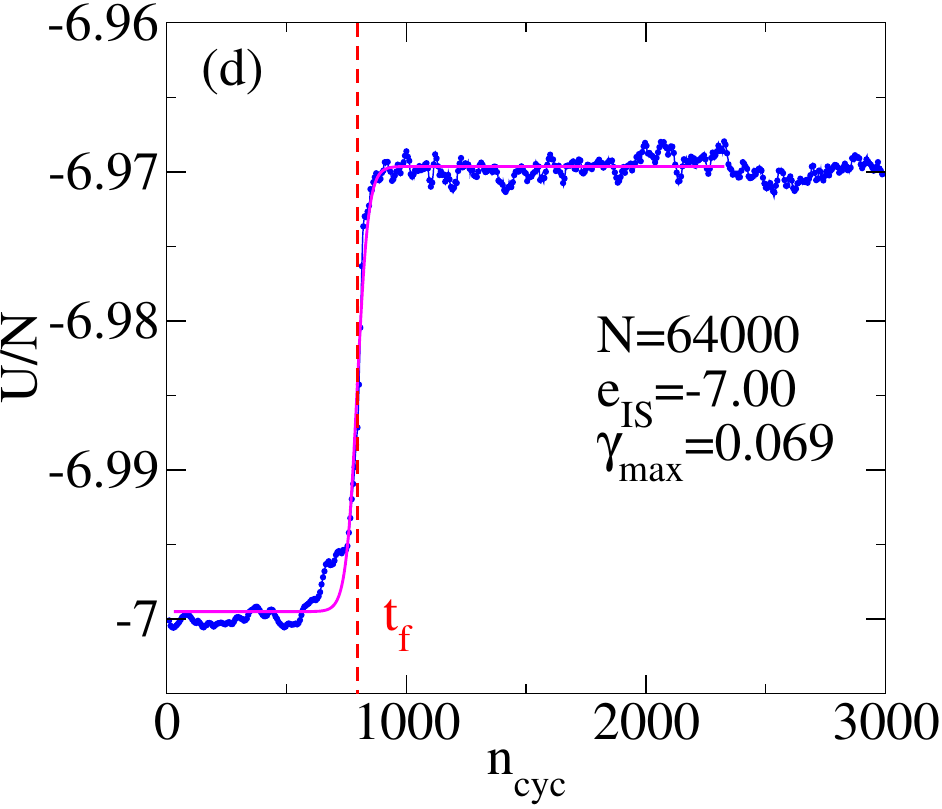}}
    \caption{ \textbf{Failure time through sigmoid fit. } Per particle potential energy $U/N$ for a single sample as a function of $n_{cyc}$ for $N=4000$ for (a) $e_{IS}=-6.96$, (b) $e_{IS}=-7.00$. Data is logarithmically binned to reduce noise.
    Analogous data for a larger system size $N=64000$ is shown (c) $e_{IS}=-6.96$, (d) $e_{IS}=-7.00$. Solid lines are fits to the sigmoid function given in the text, Eq. S3. For (c) and (d) the sigmoid function is fitted from the cycle when energy is minimum to the cycle when energy is maximum. Failure time $t_f$, obtained from the inflection point of the fit, are shown by the vertical dashed lines.}
    \label{fig:failure_time_calculation_systemsize}
\end{figure}
In the main text, we show the fitting of the sigmoid function through data for $U/N$, $q_6$, and $D^2_{min}$ for a given system size and $e_{IS}=-7.00$. Here, we explain the fitting procedure in detail, and show illustrative fits to various samples, annealing degrees, strain amplitudes, and system sizes. We employ the sigmoid function (logistic function) 
\begin{align}
    X(t)=X_{min}+\frac{X_{max}-X_{min}}{1+\exp\left(-\frac{t-t_f}{w'}\right)}\label{eq:sigmoid},
\end{align}
where $X=\in \{U/N$, $q_6$, $D^2_{min}\}$ and $t_f$ is the failure time.\\

Fig. \ref{fig:failure_time_calculation_systemsize} shows the fitting of energy data for single samples for two different system sizes $N=4000$ and $N=64000$ and two different degrees of annealing $e_{IS}=-6.96$ and $e_{IS}=-7.00$ for a strain amplitude $\gamma_{max}=0.09$.

\begin{figure}[ht!]
\flushleft \hspace{0.2cm}(a) \hspace{5.5cm} (b) \hspace{5.5cm} (c)\\
        \centering
        {\includegraphics[width=0.32\textwidth]{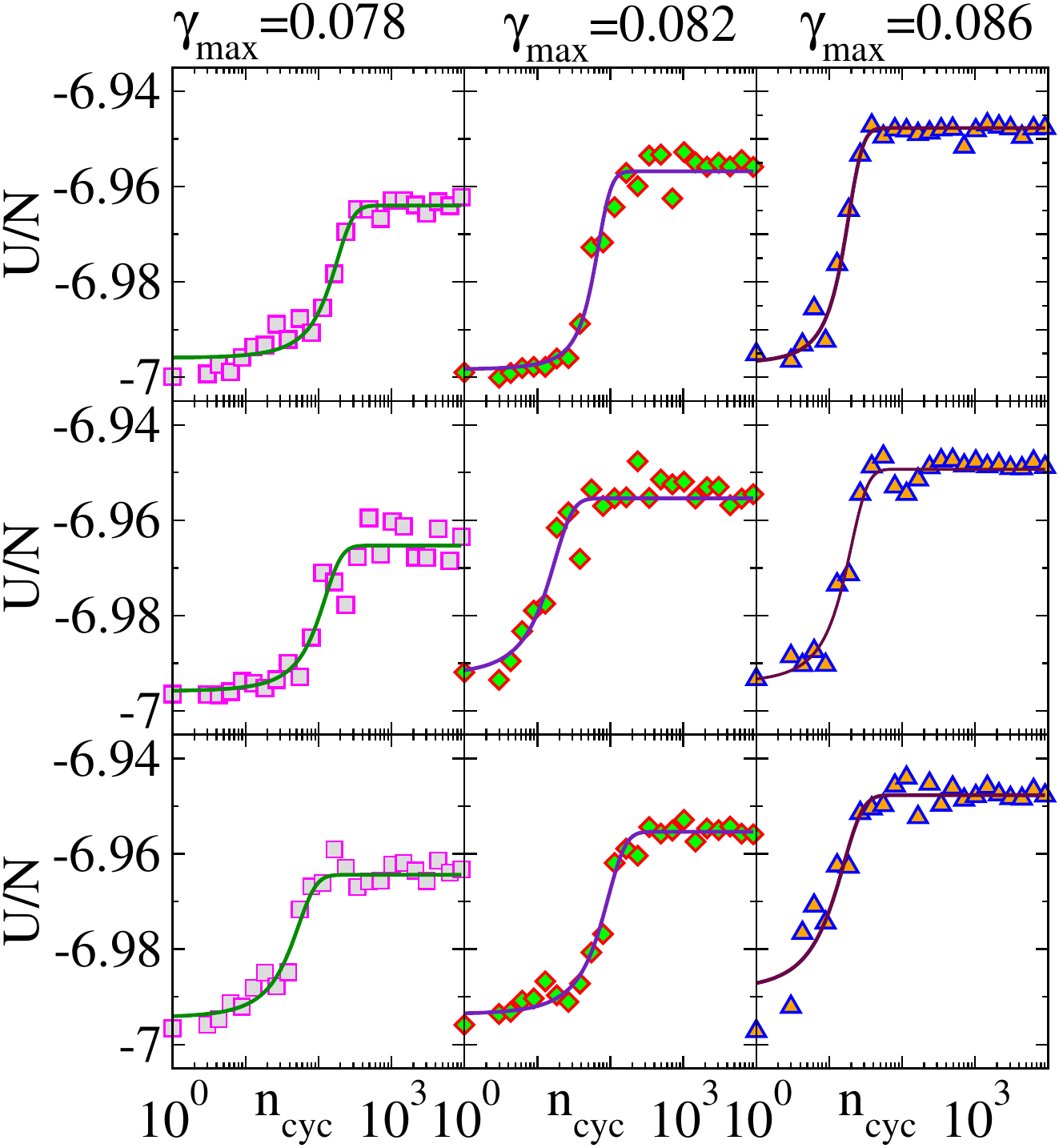}}
        {\includegraphics[width=0.32\textwidth]{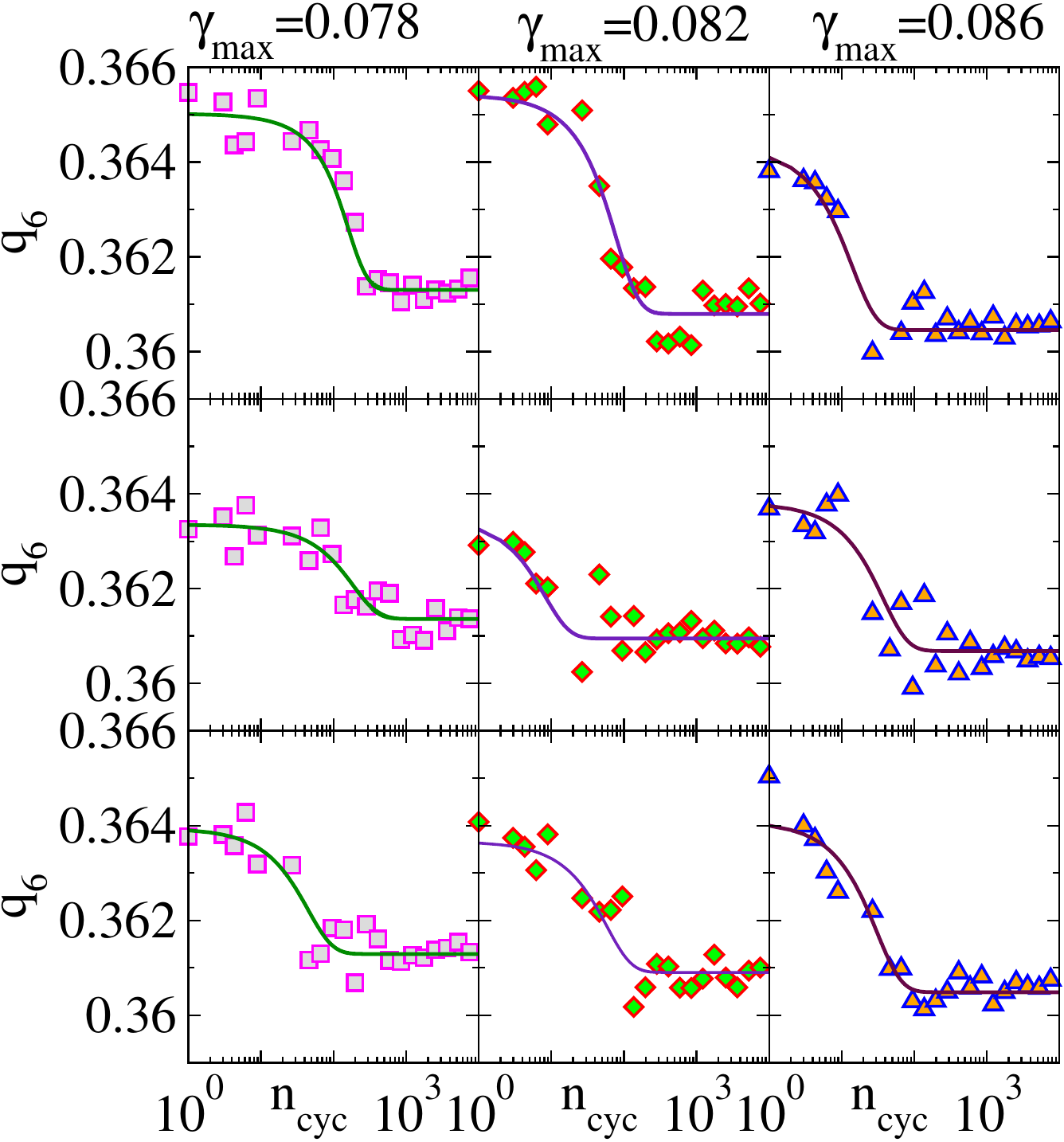}}
        {\includegraphics[width=0.32\textwidth]{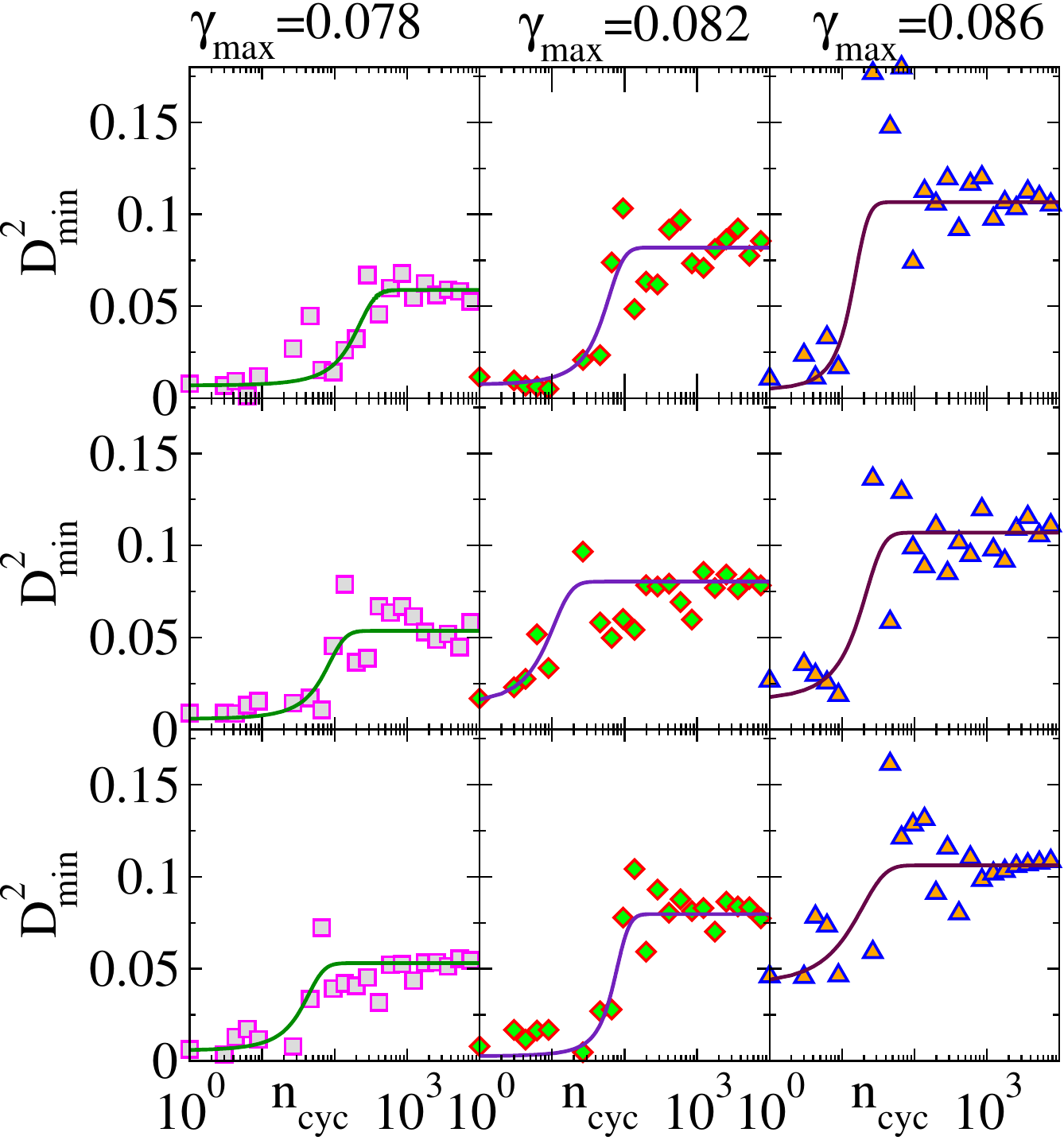}}
    \caption{ \textbf{Sigmoid fit for different quantities.} Example of sigmoid fit through the single sample data of (a) $U/N$, (b) $q_6$, (c) $D^2_{min}$ for $N=4000$ and $e_{IS}=-7.00$. Each row corresponds to the same sample, while each column represents a different strain amplitude $\gamma_{max}$.}
    \label{fig:failure_time_eIS-7.00_multiplesample_differentmetric}
\end{figure}

In Fig. \ref{fig:failure_time_eIS-7.00_multiplesample_differentmetric}, we show the sigmoid fit through data of $U/N$, $q_6$, $D^2_{min}$ for different samples and different strain amplitudes, for $N=4000$ and $e_{IS}=-7.00$. 

\begin{figure}[ht!]
\flushleft \hspace{1cm}\large{(a) \hspace{8cm} (b)}\\
        \centering
        \includegraphics[width=0.43\textwidth]{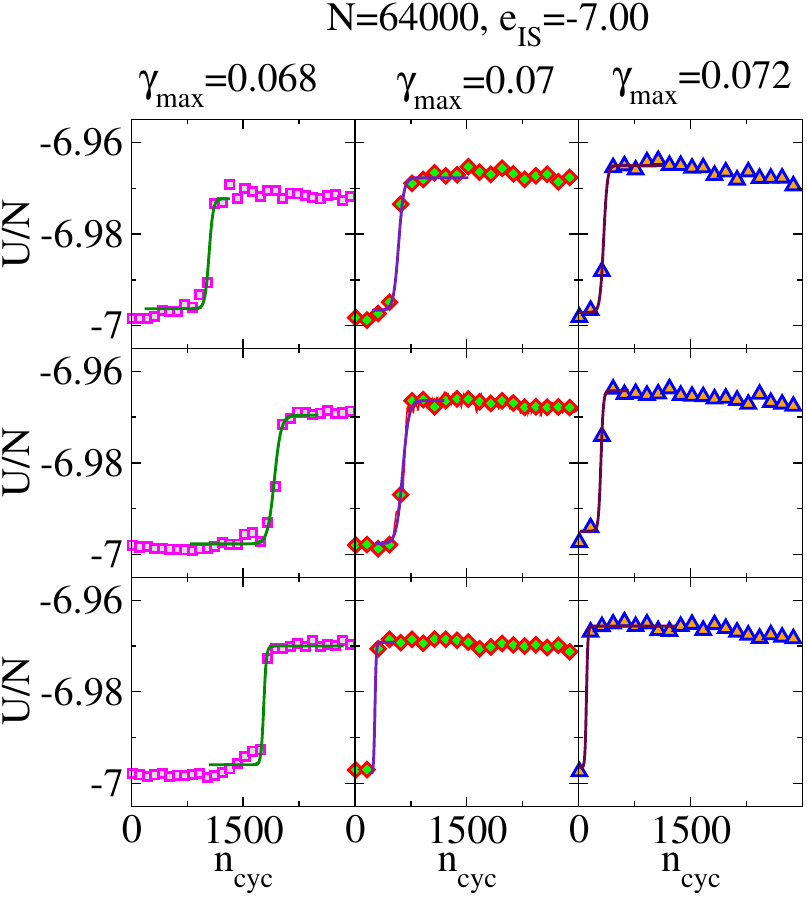}\qquad
        \includegraphics[width=0.43\textwidth]{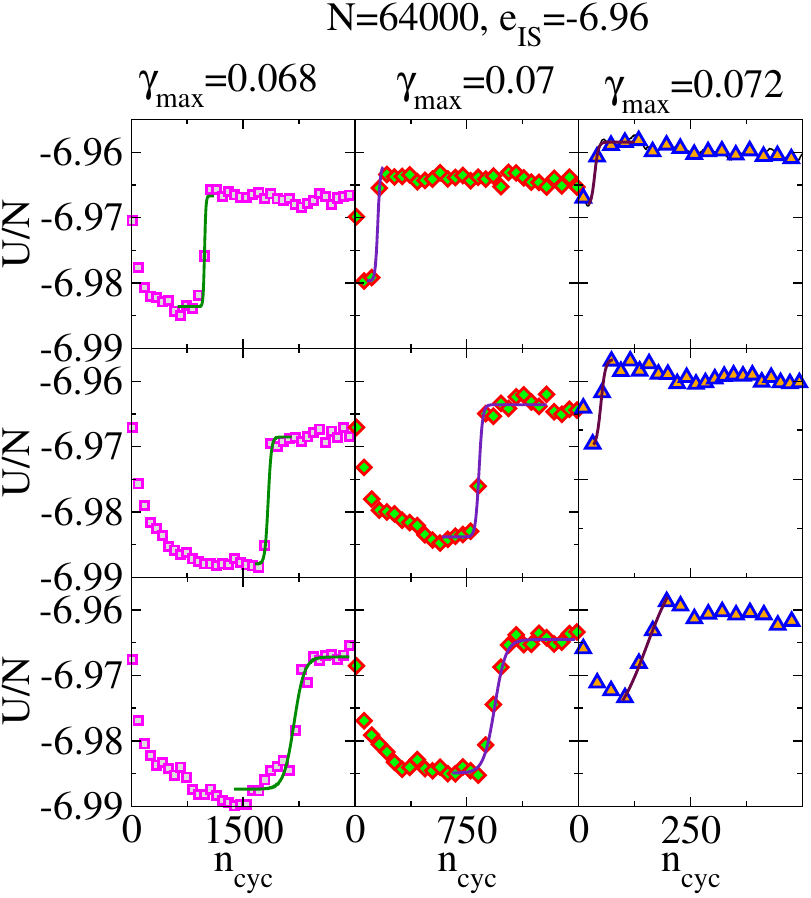}
     \caption{\textbf{Sigmoid fit for different $e_{IS}$ for large system size.} Example of sigmoid fit through single-sample data of $U/N$ with $N=64000$ for (a) $e_{IS}=-7.00$ and (b) $e_{IS}=-6.96$. Data are fitted within the range between minimum and maximum energy that contains the transition point where failure occurs.  Different rows are different samples whereas different column are different strain amplitude.}
    \label{fig:failure_time_largeSystem}
\end{figure}

In Fig. \ref{fig:failure_time_largeSystem}, we show the sigmoid fit for different samples for different strain amplitudes using $U/N$, for a large system size $N=64000$ for both $e_{IS}=-7.00$ and $e_{IS}=-6.96$. The fits describe the transition region well, and provide the failure times used in the main text.

For the poorly annealed case, of large system sizes ($N=64000, 128000$), the stroboscopic evolution of energy is non-monotonic. Initially, the energy decreases, showing annealing, and exhibits sharply defined failure after some finite number of cycles. For the well-annealed case, energy is monotonic; initially, it increases mildly with an increasing number of cycles till the system fails catastrophically. For these large system sizes, $t_f$ is calculated by fitting the sigmoid in the range between the cycles exhibiting the minimum and maximum values of energy that include the time of failure. 


For $N=4000$ and $e_{IS}=-6.96$ we consider those $\gamma_{max}$s for which the final steady state energy is greater than the initial inherent structure energy. Since for these cases, the energy profile exhibits monotonic behaviour, we fit the whole energy evolution curve to sigmoid to obtain the failure time.


\section{Evolution of stroboscopic energy with scaled number of cycles} \label{SI_PriejzevDiagram}
\begin{figure}[ht!]
       \centering
        {\includegraphics[width=.3\textwidth]{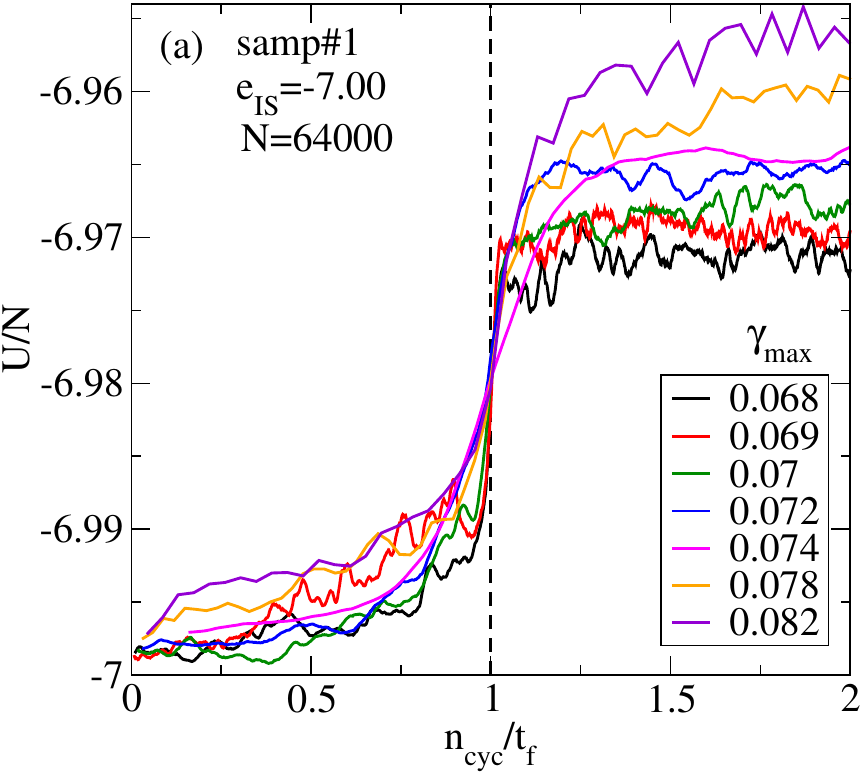}}\qquad
        {\includegraphics[width=.3\textwidth]{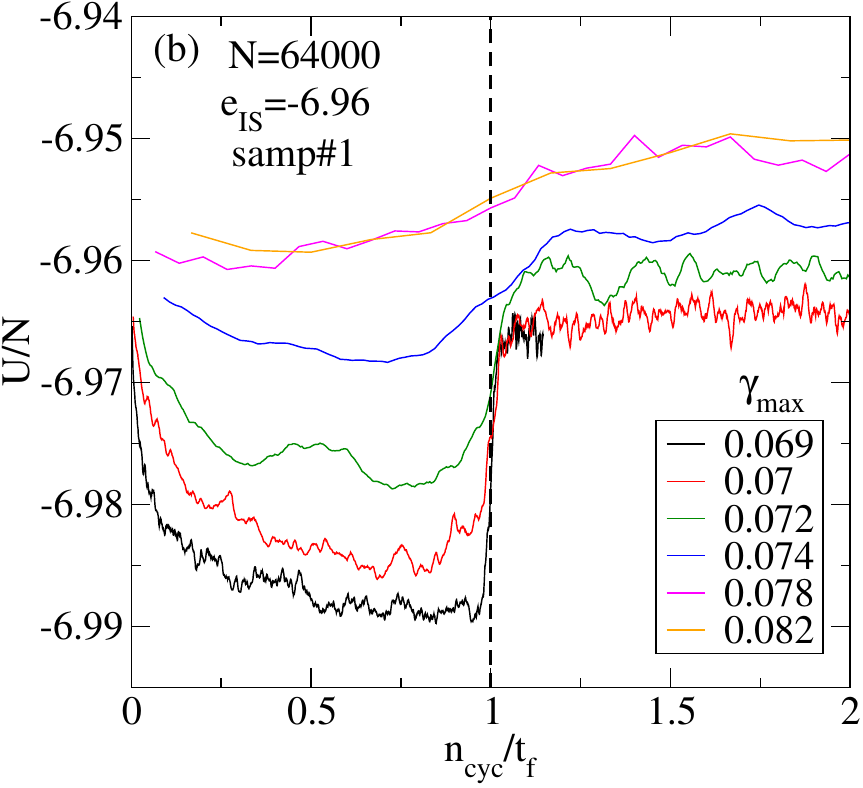}}
    \caption{ \textbf{Route to failure depends on the degree of annealing.} $U/N$ as a function of rescaled strain cycle $n_{cyc}/t_f$ for different strain amplitude $\gamma_{max}$ for (a) $e_{IS}=-7.00$ and (b) $e_{IS}=-6.96$. }
    \label{fig:Priezjev_scaling}
\end{figure}

In Fig. \ref{fig:Priezjev_scaling}, the evolution of the per-particle energy is plotted as a function of the rescaled number of cycles for different $\gamma_{max}$ for two degrees of annealing. For a well-annealed system, $e_{IS}=-7.00$, we observe that the energy follows roughly the same functional form close to  $t_f$ for a small range of $\gamma_{max}$, as also observed in Ref. \cite{PRIEZJEV2023112230}. However, for  poorly annealed glasses  ($e_{IS}=-6.96$), we do not observe such behaviour, with the energies for  $\gamma_{max}$ close to the yield value showing substantially higher annealing (reduction in energy) than for larger strain amplitudes.


\section{Calculation of failure time from the sample to sample fluctuations} \label{SI_fluct}

\begin{figure}[ht!]
    \centering
    \includegraphics[width=.98\textwidth]{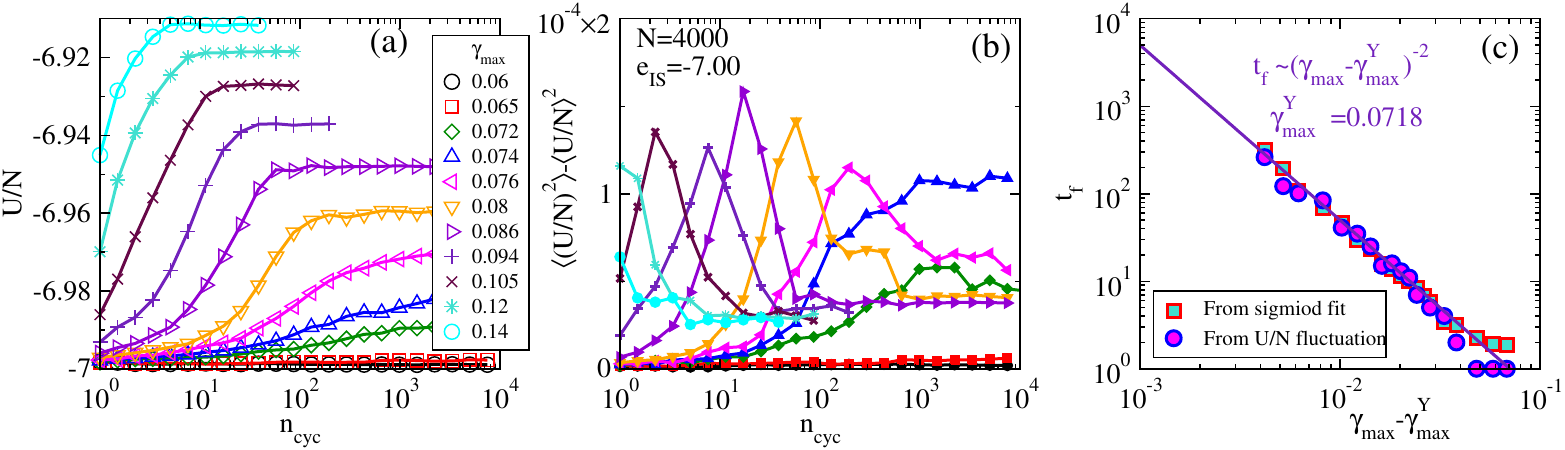}
    \caption{ \textbf{Failure time from statistical fluctuations and its scaling.}
    (a) Average per particle potential energy $U/N$ and (b) sample-to-sample fluctuation $U/N$ against $n_{cyc}$ for different $\gamma_{max}$. The failure time is determined as the cycle at which the standard deviation reaches its maximum. (c) $t_f$ obtained through this procedure is plotted as a function of $\gamma_{max}-\gamma_{max}^Y$. Data points of $t_f$ obtained from the sigmoid fit of $U/N$ (same as Fig.1(d) in main) are also compared. The solid line represents $t_f = C\left(\gamma_{max}-\gamma_{max}^Y\right)^{-\alpha}$, where the constant $C\approx 0.005$, $\gamma_{max}^Y=0.0718$, and $\alpha=2$ are the values obtained from Fig.1 of the main text.}
    \label{fig:sample_to_sample_fluctuation}
\end{figure}

The mean failure time for a strain amplitude $\gamma_{max}$ can also be obtained using the sample-to-sample fluctuations of per particle energy $U/N$. As shown in Figs. \ref{fig:sample_to_sample_fluctuation} (a) and (b), the sample-to-sample fluctuations of energy exhibit a peak close to the cycles where the sample averaged energies shows an inflection. We identify the failure time as the time for which the flucutations are maximum, as an alternate estimate.  The failure times extracted by this method also follow a power-law dependence  $t_f\sim (\gamma_{max}-\gamma_{max}^Y)^{-2}$ as shown in Fig. \ref{fig:sample_to_sample_fluctuation}(c). This comparison confirms the robustness of our estimated failure times. 

\section{Calculation of transformation width} \label{SI_width} 

We also investigate the width of the time interval, over which the failure process occurs. The quantities $X \in \{U/N, q_6, D^2_{min}\}$ can be well described by the sigmoid form as in Eq. \ref{eq:sigmoid}. For a given quantity $X$, we define the width $w$ of the transition as the  time  interval for $X$ to change from $5\%$ to $95\%$ of the range $(X_{max}-X_{min})$. The width $w$ can be calculated as follows: if at $t_1$ and $t_2$ the function $X$ reaches  $5\%$ and $95\%$ of $(X_{max}-X_{min})$ above the initial value $X_{min}$, then.

\begin{align}
    &X_1(t_1)=X_{min}+0.05(X_{max}-X_{min})\label{eq:t1}\\
                &X_2(t_2)=X_{min}+0.95(X_{max}-X_{min})\label{eq:t2}.
\end{align}
Plugging them to Eqs. \ref{eq:sigmoid}, we can have $t_1=t_f-w'\ln\left(19\right)$ and $t_2=t_f+w'\ln\left(19\right)$. Hence the width $w=2w'\ln\left(19\right)$. 


\begin{figure}[ht!]
        \centering
        \includegraphics[width=.98\textwidth]{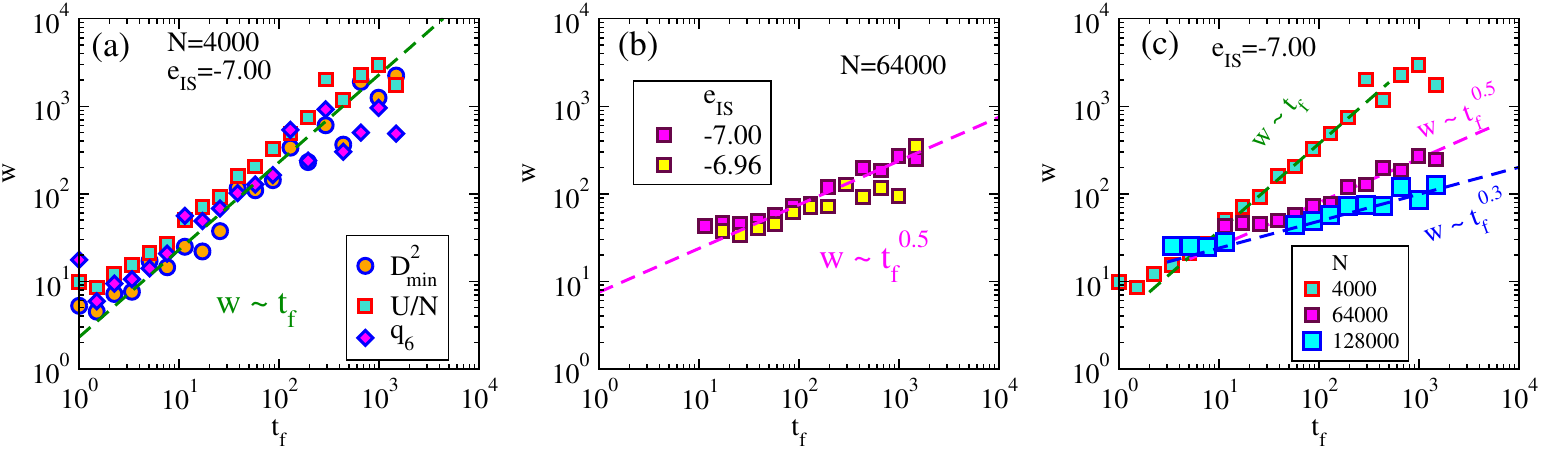}
    \caption{\textbf{Duration of failure process and its scaling with failure times.} (a) Width $w$ computed from $U/N$, $q_6$, and $D^2_{min}$, as a function of failure time $t_f$ for $N=4000$, $e_{IS}=-7.00$ . Dashed line is a fit through data points indicates $w\sim t_f$ scaling. (b) For $N=64000$, width $w$ shows a different scaling relation $w\sim t_f^{0.5}$ both for $e_{IS}=-7.00$ and $-6.96$. (c) Width {\it vs.} failure time for different system sizes and different annealing indicating continuous decrease of the exponent that relates the transformation width to the failure times, with increasing system size. In (b) and (c), $w$ is computed from $U/N$.}
    \label{fig:width_vs_failure_time_eIS-7.00_N4K_64K}
\end{figure}

In Fig. \ref{fig:width_vs_failure_time_eIS-7.00_N4K_64K}(a) we show the transformation width $w$ as a function of failure time $t_f$ for a system of size $N=4000$. Surprisingly, $w$ increases with $t_f$ linearly contrary to our expectations, but in line with the suggestion in \cite{PRIEZJEV2023112230}, where the cycle dependence of energies, when the number of cycles is scaled with the failure time, exhibited data collapse. However, when calculated  for a larger system size of ($N=64000$) as shown in Fig. \ref{fig:width_vs_failure_time_eIS-7.00_N4K_64K}(b) we find a sublinear growth of $w$ with $t_f$. To explore further, we investigate an even larger system size of ($N=128000$) and different degrees of annealing as presented in \ref{fig:width_vs_failure_time_eIS-7.00_N4K_64K}(c). We find that the power-law exponent decreases progressively with increasing system size. However, the exponent remains the same for different degrees of annealing. Based on these results, we expect that the transformation width will be a constant in the thermodynamic limit and will not depend on the failure time.

\begin{figure*}[ht!]
        \centering
         \includegraphics[width=.98\linewidth]{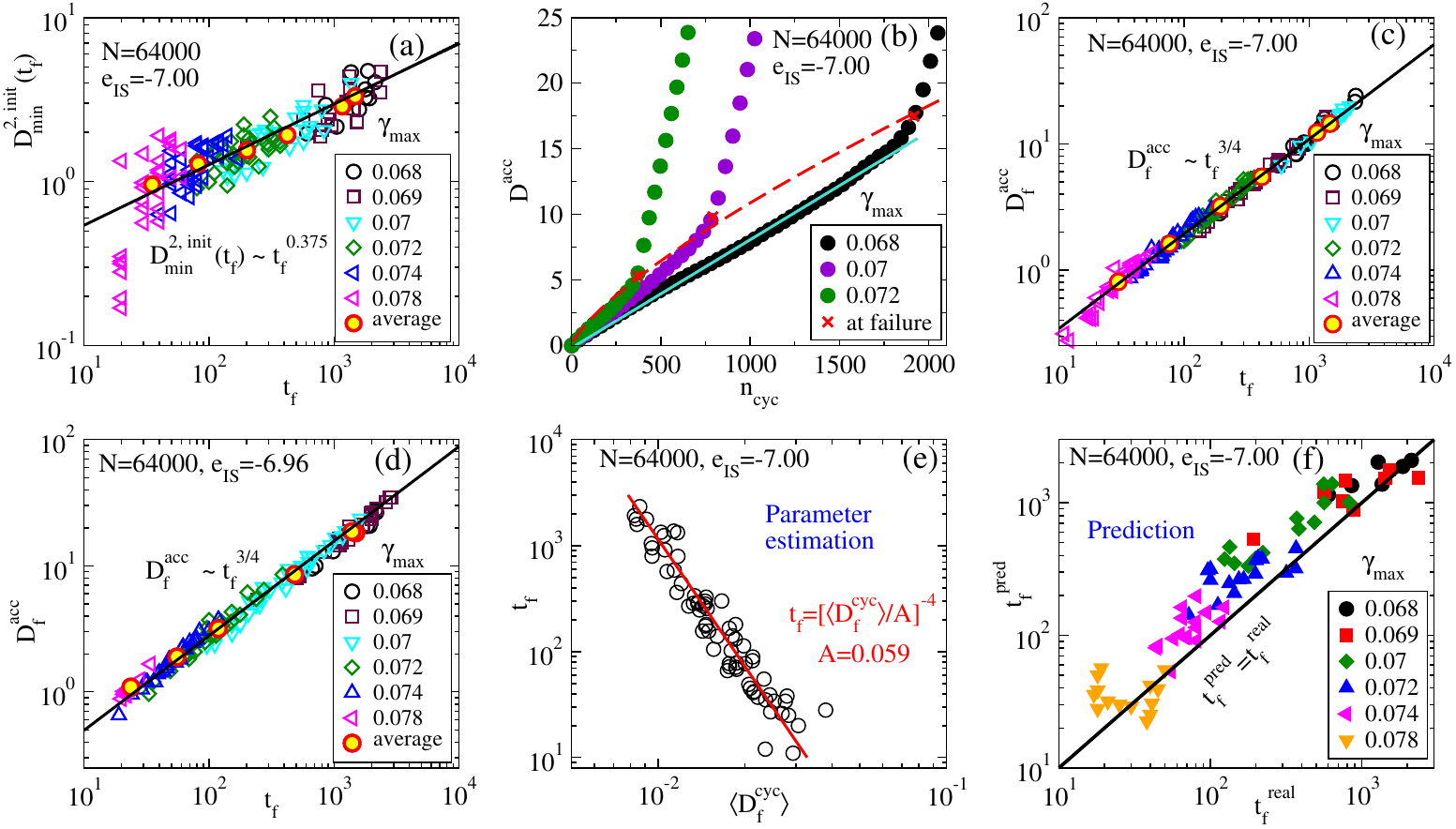}
    \caption{\textbf{Accumulated plasticity and dissipated energy up to failure time $t_f$ and predictability :} (a) Accumulated plasticity $D^{2,init}_{min}(t_{f})$, till failure, as a function of the failure time $t_{f}$ for different $\gamma_{max}$ exhibits a powerlaw with exponent $0.375$. (b) Accumulated damage $D^{acc}$ against $n_{cyc}$ for different $\gamma_{max}$. Similar to Fig.4(b) in main, here we show $D^{acc}$ equally exhibit linear growth with $n_{cyc}$ till failure $t_{f}$, beyond the initiation of failure $t_{fi}$. The crosses indicate the failure time $t_{f}$. The dashed line shows $D^{acc}_f = A t_{f}^{3/4}$. Accumulated damage till failure $D^{acc}_{f}$ grows with $t_{f}$ with a power of $3/4$ for two different degree of annealing (c) $e_{IS}=-7.00$ and (d) $e_{IS}=-6.96$. Note we observed a slightly smaller exponent than that was obtained in main Fig.4(c) with the analysis of accumulated damage up to $t_{fi}$. (e) Failure time $t_{fi}$ {\it vs.} average damage per cycle $\langle D_f^{cyc}\rangle$ (computed up to $t_{f}$) is fitted as a power law for a subset of samples to estimate the parameter $A= 0.059$ for $e_{IS}=-7.00$. (f) The predicted failure time $t_f^{pred}$ against the actual time of failure $t_{f}^{real}$ for different $\gamma_{max}$. First $n=20$ cycles are considered to compute $\langle D_n^{cyc}\rangle$ for test samples.}
    \label{figSI:damage_till_tf}
\end{figure*}

\section{Correlation of failure with accumulated plasticity and accumulated damage up to time $t_f$ }  \label{SI_damageupto_tf}

We next investigate the choice of $t_{fi}$ or $t_f$ up to which we accumulate damage (plastic activity or dissipated energy) to establish the correlation of failure time with it. Given that the failure events have finite durations in finite systems, and both plastic activity and energy dissipation undergo significant changes during the failure process, the choice of $t_{fi}$ {\it vs.} $t_f$ will lead to differences in the scaling between damage and failure times. Here we reproduce the results accumulating the damage up to $t_f$ and compare it with that reported in the main where we choose $t_{fi}$.

In Fig. \ref{figSI:damage_till_tf}(a) we show accumulated plasticity $D^{2,init}_{min}(t_{f})$, till failure, increases with $t_{f}$ as a powerlaw with exponent $0.375$. Note that while measuring accumulated plasticity up to $t_{fi}$ we found the exponent to be slightly higher, as $0.4$ [Fig 3(b) in main]. Fig. \ref{figSI:damage_till_tf}(b) shows the linear increase of accumulated damage $D^{acc}$ up to failure $t_f$. Plotting its value at failure {\it vs.} the failure time reveals $D^{acc}\sim t_f^{\beta}$ with $\beta=3/4$ as shown in Fig. \ref{figSI:damage_till_tf}(c) and (d) for a different level of annealing. Interestingly, although both exponents show slight deviations compared to the analysis in the main text, the relation of  $D^{2,init}_{min}(t_{f}) \sim t_{f}^{\beta/2}$ still holds, regardless of the choice of time up to which we accumulate damage for the analysis.

Following the same procedure described in the main text we now estimate the parameters of the correlation of failure time $t_f$ and the per-cycle damage (again computed up to $t_f$) for the set of samples used for parameter estimation, as shown in Fig. \ref{figSI:damage_till_tf}(e). We find that $t_f=[\langle D_f^{cyc}\rangle /A]^{-\beta^{\prime}}$, with $A=0.059$ and $\beta^{\prime}=4$. Notably, both the coefficient and exponent differ from the values observed in Fig. 4(e) of the main text. However, the relation $\beta^{\prime}=1/(1-\beta)$ still holds well. We further perform the prediction exercise with the new set of parameters to predict the failure time $t_f$ as shown in Fig. \ref{figSI:damage_till_tf}(f). As expected, the predictions slightly overestimate the values of the actual failure times,  likely due to the increased non-linearity around $t_f$ in the $D^{acc}$ {\it vs.} $n_{cyc}$ correlation, caused by the sharp change in energy dissipation during the failure process.

Thus, the overall results remain consistent and are not significantly affected by the choice of the time up to which damage is accumulated for analysis. Therefore, for all subsequent analyses of the poorly annealed system provided in the next section, we will use $t_f$.
\medskip

\section{Correlation of failure time with accumulated plasticity  for poorly annealed glasses} \label{SI_d2mininitpoorly}

\begin{figure*}[ht!]
        \centering
        \includegraphics[width=.98\linewidth]{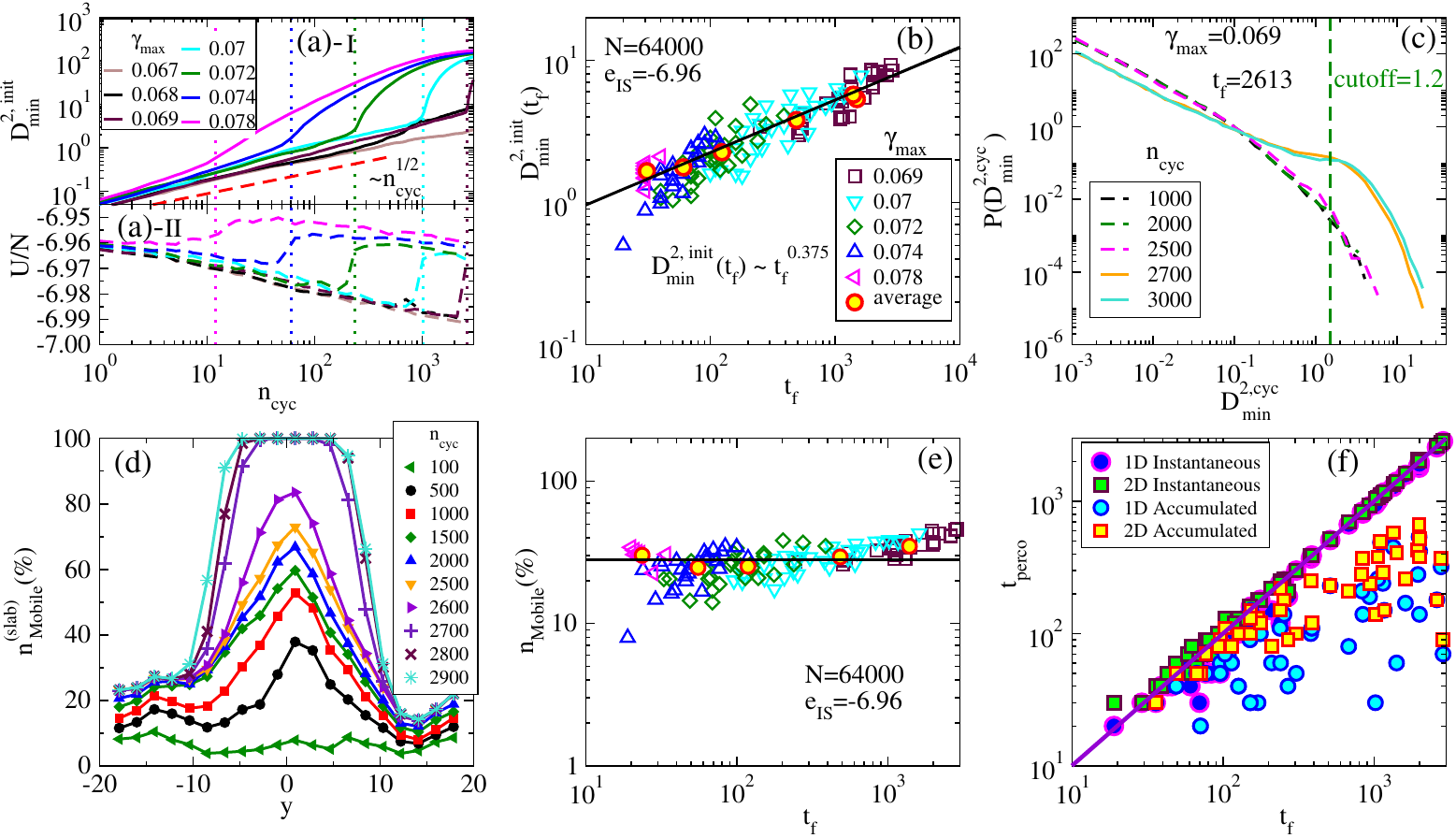}
    \caption{ \textbf{Damage as non-affine displacement for $e_{IS}=-6.96$.} 
    The non-affine displacement with respect to the initial configuration $D^{2,init}_{min}$, averaged over particles for a single sample, is shown in (a)-I (top), and the corresponding variation of the energy per particle  $U/N$ is shown in (a)-II (bottom), as a function of the number of cycles $n_{cyc}$ for $N=64000$ and $e_{IS}=-6.96$ for several $\gamma_{max}$.  The failure times are indicated with vertical dotted lines across which both the quantities increase sharply. (b) Accumulated damage till failure, $D^{2,init}_{min}(t_f)$, as a function of the failure time $t_f$ for different $\gamma_{max}$. The solid line is fit to a power law with exponent $0.375$ similar to the value for $e_{IS}=-7.00$ as shown in Fig.S8(a). (c) Distribution of cycle to cycle non-affine displacement $D^{2,cyc}_{min} \equiv D^{2}_{min} (\Delta t = 10\text{ cycle})$, computed at different numbers of cycles of strain. A threshold of $D^{2,cyc}_{min}=1.2$ is chosen beyond which the particles are identified as mobile particles. (d) Accumulation of mobile particles (for $\gamma_{max} = 0.069$) along the gradient (y) direction.  (e) The total number of mobile particles accumulated up to the failure time $t_f$ is plotted against $t_f$ for runs starting with different configurations and different $\gamma_{max}$ [(e) shares color code with (b)]. (f) The cycle at which accumulated mobile particles percolate, compared with failure time, shows that the percolation occurs close to the time of failure. 
    }
    \label{fig:D2min_from_init_PA}
\end{figure*}

We show here the results concerning the variation of accumulated plasticity as measured through $D^{2,init}_{min}$, $D^{2,cyc}_{min}$, and mobile particles and their percolation behaviour, for poorly annealed glasses, to be compared with results for well annealed glasses shown in Fig. \ref{fig:damage_accumulated_till_failure}. We present the data in Fig. \ref{fig:D2min_from_init_PA}. Compared to the well annealed case, the percolation of mobile particles happens earlier, pointing to a possible error in identifying mobile particles. Nevertheless, the other features of the correlations with failure time are as clearly present as in the well annealed case. 

\begin{figure*}[ht!]
        \centering
        \includegraphics[width=.66\linewidth]{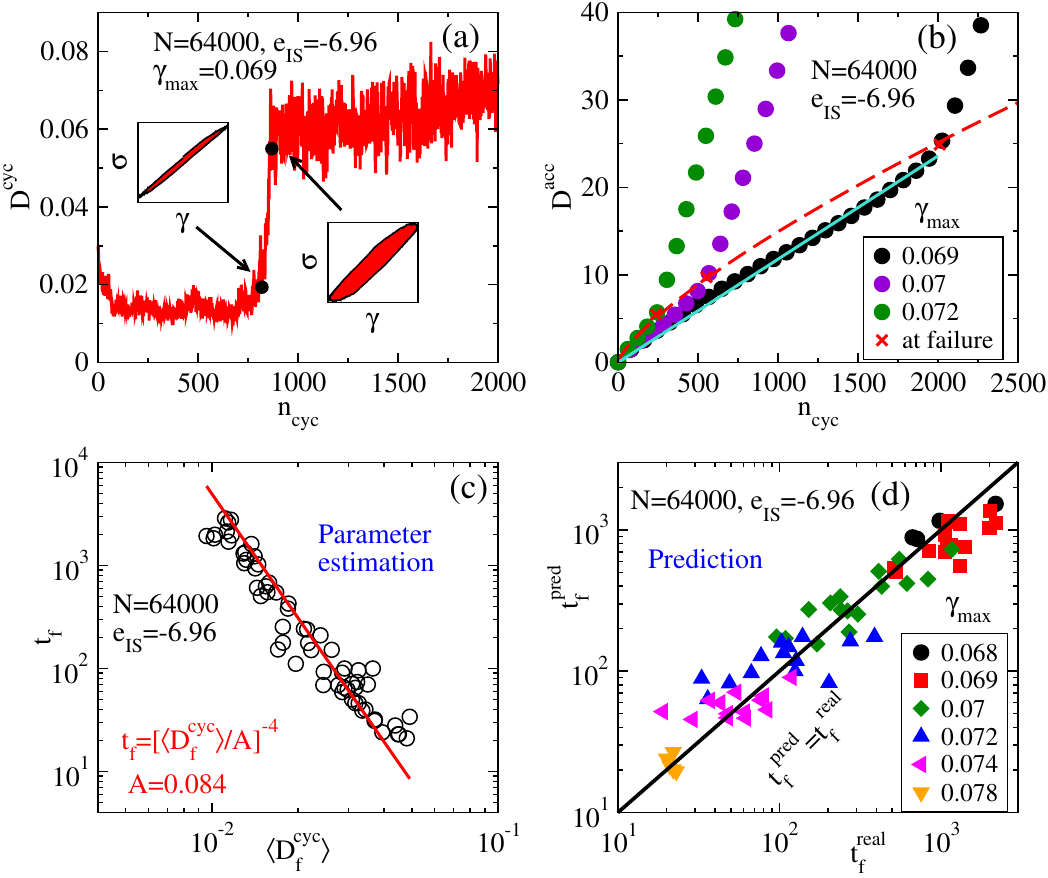} 
    \caption{\textbf{Predictability of failure times for the poorly annealed system $e_{IS}=-6.96$.} (a) Damage $D^{cyc}$ as a function of strain cycle $n_{cyc}$. The stress-strain loop at two different cycles across the failure is shown in the insets. (b) Accumulated damage against $n_{cyc}$ for different $\gamma_{max}$. All notations are the same as Fig.S8(b). Accumulated damage grows with $t_f$ as $D^{acc}_{f} = A t_{f}^{3/4}$, as shown by the dashed line. (c) For a subset of samples, the correlation of $t_{f}$ with $\langle D_f^{cyc}\rangle$ estimates the parameter $A= 0.084$ for $e_{IS}=-6.96$. (d) The predicted failure time $t_{f}^{pred}$ against the actual failure time $t_{f}^{real}$ for different $\gamma_{max}$.
    }
    \label{fig:predict_failure_EIS-6.96}
\end{figure*}

In Fig. \ref{fig:predict_failure_EIS-6.96} we show the accumulation of damage in terms of dissipative work and its correlation with $t_f$ for poorly annealed glass $E_{IS}=-6.96$. We estimate the parameters of $t_f$ {\it vs.} $\langle D_f^{cyc}\rangle$ correlation and find the exponent $\beta^{\prime}=4$ (same as well annealed case) and the prefactor $A=0.084$. As shown in \ref{fig:predict_failure_EIS-6.96}(d), the predicted failure times align well with the actual values. This indicates the robustness of predictability of failure times across different degree of annealing.

\end{document}